\documentclass[aps,preprint,showpacs, showkeys]{revtex4}%
\usepackage{amsfonts}
\usepackage{amsmath}
\usepackage{amssymb}
\usepackage{graphicx}
\usepackage{epsfig}
\usepackage{exscale}
\usepackage{float}
\usepackage{bm}
\usepackage{suffix}
\usepackage{mathtools}
\usepackage{newunicodechar}
\usepackage{bbding}
\usepackage{tikz}
\usepackage{multirow}

\providecommand{\U}[1]{\protect\rule{.1in}{.1in}}
\DeclarePairedDelimiterX\MeijerM[3]{\lparen}{\rparen}
{\begin{smallmatrix}#1 \\ #2\end{smallmatrix}\delimsize\vert\,#3}
\newcommand\MeijerG[8][]{  G^{\,#2,#3}_{#4,#5}\MeijerM[#1]{#6}{#7}{#8}}
\WithSuffix
\newcommand\MeijerG*
[7]{
G^{\,#1,#2}_{#3,#4}\MeijerM*{#5}{#6}{#7}}

\begin{document}
\title[ ]{ Precise Critical Exponents of the O(N)-Symmetric Quantum field Model using
Hypergeometric-Meijer Resummation}
\author{Abouzeid M. Shalaby}
\email{amshalab@qu.edu.qa}
\affiliation{Department of Mathematics, Statistics, and Physics, Qatar University, Al
Tarfa, Doha 2713, Qatar}

\begin{abstract}
In this work, we show that one can select different types of Hypergeometric
approximants for the resummation of divergent series with different
large-order growth factors. Being of $n!$  growth factor, the
 divergent series for the $\varepsilon$-expansion of the
critical exponents of the $O(N)$-symmetric model is approximated by the Hypergeometric
functions $_{k+1}F_{k-1}$. The divergent $_{k+1}F_{k-1}$ functions are then
resummed using their equivalent Meijer-G function representation. The convergence of the resummation results for the exponents $\nu$,\ $\eta$ and $\omega$ has been shown to improve systematically in going from low order to the highest known six-loops order. Our six-loops resummation results are  very competitive to the recent six-loops Borel with conformal mapping predictions and to recent Monte Carlo simulation results. To show that precise results extend for high $N$ values, we listed the five-loops results for $\nu$ which are  very accurate as well. The recent seven-loops order ($g$-series)  for the renormalization group functions $\beta,\gamma_{\phi^2}$ and  $\gamma_{m^2}$  have been resummed too.  Accurate predictions  for the critical coupling and the exponents $\nu$, $\eta$ and $\omega$ have been extracted from $\beta$,$\gamma_{\phi^2}$  and $\gamma_{m^2}$ approximants. 
\end{abstract}
\maketitle

\section{Introduction}

Quantum field theory (QFT) represents an important tool to study critical
phenomena for different physical systems. Critical phenomena is thus offering an indirect experimental test
to the validity of QFT. The idea stems from the universal phenomena where a
number of different systems can show up the same critical behavior in spite of 
their different microscopic details. A very clear example is the
Ising model from magnetism and the one-component  $\phi^{4}$  model from QFT
\cite{zinjustin,zin-borel,Berzin,Kleinert-Borel,kleinert,kleinert2}. The more
general example of the $\phi^{4}$ scalar field theory with $O(N)$-symmetry can
describe the critical phenomena in many physical systems that share the same respective
symmetry. Regarding the $N=0$, for example, the theory lies in the same 
universality class with polymers \cite{Polymers} while the $N=1$ case
describes the critical behavior of Ising-like models. For $N=2$, the model 
describes a preferred orientation of a magnet in a plane while the case $N=3$
can describe a rotationally invariant ferromagnet . Besides, the $N=4$ case can
mimic the phase transition in $QCD$ at finite temperature with two light
flavors \cite{QCD}.

The study of  critical phenomena within quantum field theory has been reinforced  by Wilson's introduction of the famous $\varepsilon$-expansion
\cite{Wilson,ber-wil}. Wilson ideas made the renormalization group functions to take a place in the heart of 
predicting  critical exponents from   the study of QFT models \cite{zinjustin,Berzin,Kleinert-Borel}.  However, the series generated by the $\varepsilon$-expansion is  well known to be divergent \cite{Nickel} and thus resummation techniques are indispensable   to extract reliable results from that series.  In Ref.\cite{zin-exp} (for instance),
Borel transformation with conformal mapping technique has been used to resum
divergent series of the critical exponents of the $O(N)-$symmetric model. Also
in Ref.\cite{Keleinert-st}, the five-loops $\varepsilon$-expansion of the perturbation
series for the critical exponents have been resummed using a strong-coupling
resummation technique. 

Resummation of the series generated by  $\varepsilon$-expansion has been shown
to be slightly less precise than the resummation of renormalization group
functions at fixed dimensions \cite{zin-exp}. This fact motivated the authors of the recent work 
in Ref.\cite{ON17} to move one step forward toward the improvement of
resummation predictions of the critical exponents from $\varepsilon$-expansion. In that
reference, the six-loops perturbation series of the $\varepsilon$-expansion
for the renormalization group functions of the O(N) model have been obtained and
resummed using Borel  with conformal mapping resummation algorithm.  They obtained  accurate
results for the exponents $\nu,\eta$ and $\omega$. However, this algorithm has three free parameters where their variations add  to the uncertainty in the calculations.  We will show in this work that   a simple Hypergeometric-Meijer
resummation algorithm \cite{Abo-large}, which has no free parameters,    can result in competitive  approximations
for the critical exponents from the $\varepsilon$-expansion. 

Methods that are using different approach (other than  resummation) have been used   in literature to extract accurate critical exponents of the $O(N)$ model.  Among these successful methods  is   Monte Carlo simulation which has been used to obtain accurate  critical  exponents of the $O(N)$ model \cite{MC10,MC11,nuN0E,nuN0,MCN2,MC01,MC02,MC16}. Besides, in recent years,  researchers were able to extend the applicability of conformal bootstrap methods to three dimensions which in turn resulted in very accurate predictions for the critical exponents of the $O(N)$ model too \cite{Bstrab,Bstrab5,Bstrab2,Bstrab3,Bstrab4}. The results of these techniques besides the recent Borel resummation results will be used for comparison with our predictions from Hypergeometric-Meijer resummation of divergent series representing the critical exponents.  

The divergence of perturbation series in QFT has been argued for
the first time by Dyson \cite{Dayson}. From a mathematical point of view, singularities in the complex-plane are responsible for series divergence even for small argument  \cite{singular}. The manifestation of divergence in a perturbation series appears in the form of large-order growth factors like $n!, (2n)!$ and $(3n)!$ (for instance). The appearance of such large-order behaviors  stimulates the need for resummation of such type of  perturbation series\cite{guil-res,Baker}. The most popular resummation technique is Borel and its different versions. In fact, the knowledge of the  large-order behavior of a divergent series is needed not only to accelerate the convergence of   resummation results but also to determine the type of the Borel transformation to be used. In our work, we will show that the large-order behavior is also important for   our  resummation (Hypergeometric-Meijer) algorithm \cite{Abo-large} in order to select the suitable relation between the number of numerator and denominator parameters of the used Hypergeometric approximant. 

Borel resummation and the Hypergeometric-Meijer algorithms share the need of  the large-order behavior of a divergent series to select the suitable Borel-transform and the Hypergeometric approximant respectively. There exist, however, different features for both algorithms. One can get sufficient idea about the features of  Borel resummation algorithm  by going to its extensive  use in literature.  For  the resummation of divergent series in QFT,   one can visit some  of past and recent successful studies that dealt with resummation of the divergent series of the renormalization group functions of the $O(N)$-symmetric model \cite{Kleinert-Borel,zinjustin,ON17,kleinert, zin-exp, x3-4l, zin-cr, Eta4,Guillou}. Although
resummation techniques used in literature like Borel and Borel-Pad$\acute
{e}$ can give reasonable results for the critical exponents of the $O(N)$
model, these algorithms need a relatively high order of loop calculations
which is not an easy task. To get an idea about how hard to have high orders of  loops calculations,  we assert that  it took the researchers like 25 years to move forward from five-loops to six-loops calculations \cite{Kleinert5L,ON17}. Even at the level of more simpler
theories like the $\mathcal{PT-}$symmetric $i\phi^{3}$ field theory, the four
loops renormalization group functions have been just recently obtained
\cite{x3-4l}. In going to more complicated theories that have fermionic as
well as gauge boson sectors, the calculation of a relatively high loop orders
is not an easy task. The Hypergeometric-Meijer algorithm, on the other hand,  can give reasonable results even in using few orders  from a perturbation series as input. It is thus very suitable for the  study of non-perturbative features of a quantum field theory.

In   Borel algorithms, results are always achieved via numerical calculations. This feature leads to the resummation of  individual physical amplitudes one by one.  The existence of  a resummation algorithm that avoids this feature might help in getting  other amplitudes without further resummation steps. Instead, we can obtain them from simple calculus. For instance, the vacuum energy or equivalently the effective potential is
known to be the generating functional of the one-particle-irreducible
amplitudes. Accordingly, getting a closed form resummation function  for the effective potential enables  one to get other amplitudes via functional
differentiation \cite{Peskin,Abo-exact}. The Hypergeometric-Meijer resummation as we
will see can give accurate results as well as being simple and of closed form.
Besides, it does not  have any free parameters to fix like other resummation
algorithms which use optimization tools  to fix the introduced free parameters.

The Hypergeometric-Meijer resummation algorithm we use in this work is a
development of the recently introduced simple  Hypergeometric resummation
algorithm \cite{Prl}. In the Hypergeometric algorithm, the
Hypergeometric approximant $_{2}F_{1}(a,b; c;\sigma z)$ has been suggested for
the resummation of a divergent series. The four parameters $a,b,c$ and
$\sigma$ are \ obtained by comparing the first four orders of the expansion of
$_{2}F_{1}(a,b;c;\sigma z)$ in the variable $z$ with the four available orders
of the divergent series under consideration. To illustrate this more, consider
a series representing a physical quantity $Q\left(  z\right)  $ as:%
\begin{equation}
Q\left(  z\right)  =\sum_{0}^{4}c_{i}z^{i}+O\left(  z^{5}\right)  ,
\end{equation}
we have also the series expansion of $c_{0}$\ $_{2}F_{1}(a,b; c; \sigma z)$ as:%
\begin{align}
c_{0}\ _{2}F_{1}(a,b; c; \sigma z)  &  =c_{0}+c_{0}\frac{ab\sigma}{c}%
z+c_{0}\frac{a(a+1)b(b+1) \sigma^{2}}{2c(c+1)}z^{2}\nonumber\\
&  +c_{0}\frac{a(a+1)(a+2)b(b+1)(b+2)  \sigma^{3}}{6c(c+1)(c+2)}%
z^{3}\\
&  +c_{0}\frac{a(a+1)(a+2)(a+3)b(b+1)(b+2)(b+3)  \sigma^{4}%
}{24c(c+1)(c+2)(c+3)}z^{4}\nonumber\\
&  +..............\nonumber
\end{align}

For $c_{0}\ _{2}F_{1}(a,b; c;\sigma z)$ to serve as an approximant for
$Q\left(  x\right)  $ we have to set%

\begin{align}
c_{1}  &  =\ c_{0}\frac{ab\sigma}{c}\nonumber\\
c_{2}  &  =c_{0}\frac{a(a+1)b(b+1)c\sigma^{2}}{2c(c+1)}\nonumber\\
c_{3}  &  =c_{0}\frac{a(a+1)(a+2)b(b+1)(b+2) \sigma^{3}%
}{6c(c+1)(c+2)} \label{eqset}\\ 
c_{4}  &  =c_{0}\frac{a(a+1)(a+2)(a+3)b(b+1)(b+2)(b+3) \sigma
^{4}}{24c(c+1)(c+2)(c+3)}\nonumber, %
\end{align}
which can be solved to determine the unknown parameters $a,b,c,d,\sigma$ in
terms of the known coefficients $c_{1},c_{2},c_{3}$ and $c_{4}$.

To accelerate the convergence of the algorithm, we suggested the  employment of 
parameters from the asymptotic behavior of the perturbation series at large
values of the argument $z$ \cite{Abo-hyper} or equivalently the strong
coupling data. Our suggestion is based on the realization that when $a-b$ is
not an integer, the Hypergeometric function has the following asymptotic form
\cite{HTF};%
\[
_{2}F_{1}\left(  a,b; c ;g\right)  \sim\lambda_{1}g^{-a}+\lambda_{2} g^{-b},\left\vert g\right\vert \gg1.
\]
Also the method has been generalized to accommodate higher orders from the
perturbation series by using the generalized Hypergeometric function $_{\text{
}p}F_{p-1}(a_{1},...a_{p}; b_{1}....b_{p-1}; \sigma z)$ where the $a_{i}$
parameters are extracted  from the asymptotic behavior of the perturbation
series at large $z$ value.

The Hypergeometric algorithm either the version in Ref.\cite{Prl} or
Ref.\cite{Abo-hyper} cannot accommodate the large order data available for
many perturbation series in physics. The point is that the series expansion of the  Hypergeometric
function  $_{2}F_{1}(a,b; c; \sigma z)$ has a finite radius of
convergence while it has been used for the resummation of a divergent series with zero radius of convergence. This means that the large order behavior of the expansion of  the function $_{2}F_{1}(a,b; c; \sigma z)$  can not account explicitly for the $n!$ growth factor
characterizing  a perturbation series with zero radius of convergence. In fact, in the Hypergeometric algorithm, the parameter
$\sigma$ ought to take large values to compensate for that \cite{cut,Prd-GF}
but itself cannot be considered as a large-order parameter. Indeed, employing
parameters from large-order behavior is well known to accelerate the
convergence of resummation algorithms (Borel for instance). Moreover, one can
not apply the suitable Borel transform (divide  by $n!$ for instance ) unless we know the
large order behavior of the perturbation series. These facts led us to develop
the Hypergeometric algorithm \cite{Abo-large} by using the approximants
$_{p}F_{p-2}(a_{1},a_{2},....,a_{p}; b_{1},b_{2},....b_{p-2}; \sigma z)$ instead
of $_{2}F_{1}(a,b; c;\sigma z)$. The Hypergeometric functions $_{p}%
F_{p-2}(a_{1},a_{2},....,a_{p}; b_{1},b_{2},....b_{p-2}; \sigma z)$ are all
sharing the same analytic properties (with respect to $z$) and all have
expansions of zero-radius of convergence as well as having an
$n!$ growth factor. Possessing the main features of the divergent series under consideration, the Hypergeometric function $_{p}F_{p-2}(a_{1},a_{2},....,a_{p}; b_{1},b_{2},....b_{p-2}; \sigma z)$  is thus an ideal candidate  for the resummation of that series.

The structure of the paper is as follows. In sec.\ref{Hyper},
we introduce the generalized Hypergeometric-Meijer  algorithm for the resummation
of a divergent series with a growth factor of the form  $((p-q-1)n)!$. In sec.\ref{appl}, we use the algorithm to resum the $\varepsilon
-$expansions of the exponents $\nu(\nu^{-1}),\eta$ and $\omega$ and the critical coupling
up to five-loops of the $O(N)$-symmetric model. The resummation results for the recent six-loops order is presented for the exponents
$\nu (\nu^{-1}),\eta$ and $\omega$ in sec.\ref{sixL}. Resummation of the seven-loops of the
$g-$expansion of the renormalization group functions,   which   has no resummation trials  in literature so far,  is presented in sec.\ref{sevenL}. Summary and conclusions will follow in sec.\ref{conc}.

\section{The generalized Hypergeometric-Meijer Resummation algorithm}
\label{Hyper}
Consider a divergent series that represents a physical amplitude $Q(z)$ as
\begin{equation}
Q\left(  z\right)  =\sum_{n=0}^{M}c_{n}z^{n}+O\left( z^{ M+1}\right),
\end{equation}
where the first $M+1$ orders are known. Assume that  the large-order behavior of that series takes the from:
\begin{equation}
c_{n}\sim\alpha n!(-\sigma)^{n}n^{b}\left(  1+O\left(  \frac{1}{n}\right)
\right)  ,\text{ \ \ }n\rightarrow\infty. \label{LOB}%
\end{equation}
In Ref.\cite{Abo-large}, we showed that  when $p=q+2$, the perturbative expansion of the
Hypergeometric function $_{\text{ }p}F_{q}(a_{1},...a_{p};b_{1}....b_{q}%
; -\sigma z)$ which has a zero-radius of convergence  can be parametrized to
give the same large-order behavior of the above perturbation series.
Accordingly, one sets the constraint $\sum_{i=1}^{p}a_{i}-\sum_{i=1}^{p-2}%
b_{i}-2=b$, besides the constraints set by matching the perturbation expansion
of \ $_{\text{ }p}F_{q}(a_{1},...a_{p};b_{1}....b_{q};-\sigma z)$ \ with the
available orders of the divergent series. Then the parametrized divergent
series of $_{\text{ }p}F_{q}(a_{1},...a_{p}; b_{1}....b_{q}; \sigma z)$ is
resummed using its representation in terms of Meijer-G function as follows
\cite{HTF}:
\begin{equation}
_{\text{ }p}F_{q}(a_{1},...a_{p};b_{1}....b_{q};z)=\frac{\prod_{k=1}^{q}%
\Gamma\left(  b_{k}\right)  }{\prod_{k=1}^{p}\Gamma\left(  a_{k}\right)  }%
\MeijerG*{1}{p}{p}{q+1}{1-a_{1}, \dots,1-a_{p}}{0,1-b_{1}, \dots, 1-b_{q}}{z}.
\label{hyp-G-C}%
\end{equation}
Note that the authors in Ref.\cite{Prd-GF} used a Borel-Pad$\acute{e}$
algorithm that leads to Meijer-G approximants parametrized by weak-coupling information.

One can generalize the idea of our previous work in Ref.\cite{Abo-large} to
other types of divergent series with growth factors other
than $n!$. For instance, the divergent series of the ground state energy of
the sixtic anharmonic oscillator has a zero radius of convergence but the
growth factor is $(2 n)!$ while it is $(3n)!$ for the octic anharmonic
oscillator \cite{x6-x8}. Knowing that the asymptotic form of the ratio of two $\Gamma$ functions is given by \cite{Gamma}:%

\begin{equation}
\frac{\Gamma\left(  n+\alpha\right)  }{\Gamma\left(  n+\beta\right)
}=n^{\alpha-\beta}\left(  1+\frac{\left(  \alpha-\beta\right)  \left(
-1+\alpha+\beta\right)  }{n}+O\left(  \frac{1}{n^{2}}\right)  \right),
\end{equation}
one can easily conclude that either the Hypergeometric approximants $\protect_{\text{
}p}F_{p-1}(a_{1},...a_{p}; b_{1}....b_{p-1}; \sigma z)$ used in
Ref.\cite{Abo-hyper} or $_{\text{ }p}F_{p-2}(a_{1},...a_{p}; b_{1}%
....b_{p-2}; \sigma z)$ used in Ref.\cite{Abo-large} cannot account for the
growth factors of the sixtic or octic ground state energies. Accordingly, one
can accept that there exists more than one type of Hypergeometric functions (different $S=p-q$)
that are needed to approximate different divergent series in physics with different large-order growth factors. 

Based on the idea that   the large-order asymptotic behavior is responsible for the selection of  the suitable  Hypergeometric approximant for a perturbation series, one can list  different   $_{\text{ }p}F_{q}(a_{1},...a_{p}; b_{1}....b_{q}; -\sigma z)$ approximants for different growth factors as follows:

\begin{enumerate}
\item for divergent series that has the large-order behavior in Eq.(\ref{LOB})
($n!$ growth factor), the suitable resummation function is $_{\text{ }p}%
F_{p-2}(a_{1},...a_{p}; b_{1}....b_{p-2}; \sigma z)$.

\item For a series that has a large-order behavior like $ \gamma
\Gamma\left(  2n+\frac{1}{2}\right)  \ (-\sigma)^{n}n^{b}\ ,$
\ \ $n\rightarrow\infty$, the suitable one is $_{\text{ }p}F_{p-3}%
(a_{1},...a_{p}; b_{1}....b_{p-3}; -\sigma z)$. This is because one
can easily show that for $p=q+3$, one can get a similar large-order
behavior. An example of such divergent series is   the ground state
energy of the sixtic anharmonic oscillator \cite{x6-x8}

\item For the ground state energy of the octic anharmonic oscillator, the large
order behavior is given by $ \sim\delta\ \Gamma\left(  3n+\frac{1}%
{2}\right)  \ (-\sigma)^{n}n^{b}\ ,$ \ \ $n\rightarrow\infty$, which can be
reproduced by the generalized Hypergeometric function $_{\text{ }p}F_{p-4}%
(a_{1},...a_{p}; b_{1}....b_{p-4}; -\sigma z)$.

\item For a divergent series that has a finite radius of convergence, the
suitable resummation function is $_{\text{ }p}F_{p-1}(a_{1},...a_{p}%
; b_{1}....b_{p-1}; \sigma z)$. An example of such series is the
ground state energy of the Yang-Lee model (Eq.(86) in Ref.\cite{zin-borel}).
\end{enumerate}
Based on this classification,   knowing the large order behavior of a divergent
series is essential not only to accelerate the convergence of the resummation
algorithm but also to determine the suitable Hypergeometric approximant. A
note to be mentioned is that, for $p\geq q+2$, the Hypergeometric function
$_{\text{ }p}F_{q}(a_{1},...a_{p}; b_{1}....b_{q}; \sigma z)$ has a zero radius
of convergence but it can be resumed using the closely related Meijer-G
function (see Eq.(\ref{hyp-G-C})) which has the integral form \cite{HTF}:
\begin{equation}
\MeijerG*{m}{n}{p}{q}{c_{1}, \dots,c_{p}}{d_{1}, \dots, d_{q}}{z} =\frac{1}{2\pi
i}\int_{C}\frac{\prod_{k=1}^{n}\Gamma\left(  s-c_{k}+1\right)  \prod_{k=1}%
^{m}\Gamma\left(  d_{k}-s\right)  }{\prod_{k=n+1}^{p}\Gamma\left(
-s+c_{k}\right)  \prod_{k=m+1}^{q}\Gamma\left(  s-d_{k}+1\right)  }z^{s}ds.
\end{equation}
The Hypergeometric-Meijer algorithm which will be used in this work to resum the divergent series representing the critical exponents of the
$O(N)$ vector model can be thus summarized in two simple steps \cite{Abo-large}:

\begin{enumerate}
\item Parametrize the Hypergeometric function $_{\text{ }p}F_{p-2}%
(a_{1},...a_{p};b_{1}....b_{p-2};\sigma z)$ using both weak-coupling and
large-order data of the series under consideration (for   $\varepsilon-$expansion, the strong coupling data
  represented by the numerator parameters $a_{i}$  is not  known yet).

\item Resum the divergent $_{\text{ }p}F_{p-2}(a_{1},...a_{p};b_{1}%
....b_{p-2};\sigma z)$ function using the representation in terms of the
Meijer-G function in Eq.(\ref{hyp-G-C}).
\end{enumerate}

There exist some technical issues when applying the algorithm. The first issue
is that for high orders, computer can take a relatively long-time to solve the
set of equations like the one in Eq.(\ref{eqset}). To overcome this problem,
we generated the ratio $R_{n}=\frac{c_{n}}{c_{n}-1}$ and then solve the set of
equations:%
\begin{equation}
R_{n}=\frac{1}{n}\frac{%
{\displaystyle\prod_{i=1}^{p}}
\left(  a_{i}+n-1\right)  }{%
{\displaystyle\prod_{j=1}^{q}}
\left(  b_{j}+n-1\right)  }\sigma.
\end{equation}
For example,    the approximant  $_{\text{ }p}F_{q}(a_{1},...a_{p};b_{1}....b_{q};\sigma
z)$ generates the following  set of equations:%
\begin{align}
R_{1}  &  =\frac{a_{1}a_{2} ............a_{p}}{b_{1}b_{2} ............b_{q}}\sigma \nonumber\\
R_{2}  &  =\frac{\left(  a_{1}+1\right)  \left(  a_{2}+1\right)   ............\left(
a_{p}+1\right)  }{2\left(  b_{1}+1\right)  ............\left(  b_{q}+1\right)
}\sigma \nonumber\\
&  .\\
&  .\nonumber\\
&  .\nonumber\\
R_{p+q}  &  =\frac{\left(  a_{1}+p+q-1\right)   ............\left(  a_{p}+p+q-1\right)
}{\left(  p+q\right)  \left(  b_{1}+p+q-1\right)   ............\left(
b_{q}+p+q-1\right)  }\sigma \nonumber.
\end{align}
This trick decreases the degree of non-linearity in the set of equations and
thus saves the computational time.

The other issue regarding the application of the Hypergeometric-Meijer
algorithm is that at some orders one might  find no solution for the set of
equations defining the parameters in the Hypergeometric function. In this case,
one resorts to a successive subtraction of the perturbation series. This trick is
well known in resummation algorithms \cite{Kleinert-Borel,Prd-GF}. However,
the subtracted series will have a different large-order $b$ parameter where it
increases by one per each subtraction ( see for instance sec.16.6 in
Ref. \cite{Kleinert-Borel}).

\section{Hypergeometric-Meijer Resummation for the $\varepsilon-$ expansion of
 critical exponents and coupling  up to five loops \label{appl}}

\label{5L}

The Lagrangian density of the $O(N)$-vector model is given by:%
\begin{equation}
\mathcal{L=}\frac{1}{2}\left(  \partial\Phi\right)  ^{2}+\frac{m^{2}}{2}%
\Phi^{2}+\frac{\lambda}{4!}\Phi^{4},
\end{equation}
where $\Phi=\left(  \phi_{1},\phi_{2},\phi_{3},...........\phi_{N}\right)  $
is an N-component field with $O(N)$ symmetry such that $\Phi^{4}=\left(
\phi_{1}^{2}+\phi_{2}^{2}+\phi_{3}^{2}+...........\phi_{N}^{2}\right)  ^{2}$.
At the fixed point, the $\beta$-function is zero which sets a critical
coupling as a function of $\varepsilon=4-d.$ Accordingly, one can
obtain the renormalization group functions as power series in $\varepsilon$.
In  the following parts of this section, we list the resummation results (up to five loops) for the
exponents $\nu, \eta$ and $\omega$ as well as the critical coupling  of that model.

\subsection{ Two, three, four and five loops resummation for the exponent
$\nu$}

Up to five-loops, the power series for the reciprocal of the critical exponent
$\nu$ is given by \cite{Kleinert-Borel}:
\begin{equation}
\nu^{-1}\approx 2+\sum_{i=1}^{5}c_{i}\varepsilon^{i}, \label{eps-pert}%
\end{equation}
where
\begin{align}\label{ccoef}
c_{1} &  =\frac{N+2}{N+8}\nonumber\\
c_{2} &  =-\frac{(N+2)(13N+44)}{2(N+8)^{3}}\nonumber\\
c_{3} &  =\frac{(N+2)}{8(N+8)^{5}}\{3N^{3}-452N^{2}+96(N+8)(5N+22)\zeta
(3)-2672N-5312\}\nonumber\\
c_{4} &  =\frac{(N+2)}{32(N+8)^{7}}\{3N^{5}+398N^{4}-12900N^{3}-1280(N+8)^{2}%
\left(  2N^{2}+55N+186\right)  \zeta(5)\nonumber\\
&  \text{ \ \ \ \ \ \ \ \ \ \ \ \ \ \ \ \ \ \ \ \ \ \ \ \ \ }+16(N+8)\left(
3N^{4}-194N^{3}+148N^{2}+9472N+19488\right)  \zeta(3)\nonumber\\
&  \text{ \ \ \ \ \ \ \ \ \ \ \ \ \ \ \ \ \ \ \ \ \ \ \ \ \ }-81552N^{2}%
-219968N+\frac{16}{5}\pi^{4}(N+8)^{3}(5N+22)-357120\}\nonumber\\
c_{5} &  =\frac{(N+2)}{128(N+8)^{9}}\{3N^{7}-1198N^{6}-27484N^{5}%
-1055344N^{4}-5242112N^{3}\nonumber\\
&  \text{ \ \ \ \ \ \ \ \ \ \ \ \ \ \ \ \ \ \ \ \ \ \ \ \ }-5256704N^{2}%
+56448(N+8)^{3}\left(  14N^{2}+189N+526\right)  \zeta(7)\nonumber\\
&  \text{ \ \ \ \ \ \ \ \ \ \ \ \ \ \ \ \ \ \ \ \ \ \ \ \ }%
+6999040N-626688-\frac{1280}{189}\pi^{6}(N+8)^{4}\left(  2N^{2}%
+55N+186\right)  \nonumber\\
&  \text{ \ \ \ \ \ \ \ \ \ \ \ \ \ \ \ \ \ \ \ \ \ \ \ \ }+256(N+8)^{2}%
\zeta(5)\left(  155N^{4}+3026N^{3}+989N^{2}-66018N-130608\right)  \nonumber\\
&  \text{ \ \ \ \ \ \ \ \ \ \ \ \ \ \ \ \ \ \ \ \ \ \ \ \ }-1024(N+8)^{2}%
\left(  2N^{4}+18N^{3}+981N^{2}+6994N+11688\right)  \zeta(3)^{2}\nonumber\\
&  \text{ \ \ \ \ \ \ \ \ \ \ \ \ \ \ \ \ \ \ \ \ \ \ \ \ }+\frac{8}{15}%
\pi^{4}(N+8)^{3}\left(  3N^{4}-194N^{3}+148N^{2}+9472N+19488\right)
\nonumber\\
&  \text{ \ \ \ \ \ \ \ \ \ \ \ \ \ \ \ \ \ \ \ \ \ \ \ \ }-16(N+8)\zeta
(3)[13N^{6}-310N^{5}+19004N^{4}+102400N^{3}-381536N^{2}\nonumber\\
&  \text{ \ \ \ \ \ \ \ \ \ \ \ \ \ \ \ \ \ \ \ \ \ \ \ \ }%
-2792576N-4240640]\}.
\end{align}
The large-order parameters takes the form in Eq.(\ref{LOB}) where
\cite{Kleinert-Borel}
\[
\sigma=\frac{3}{N+8}\text{ and \ }b=4+\frac{N}{2}.
\]
The suitable Hypergeometric  approximant is thus 
 $_{\text{ }p}F_{p-2}(a_{1},...a_{p}; b_{1}....b_{p-2}; -\sigma z)$ where it 
can reproduce \ the large order behavior in Eq.(\ref{LOB}).
The number of unknown parameters in $_{\text{ }p}F_{p-2}(a_{1},...a_{p}%
; b_{1}....b_{p-2}; -\sigma z)$ is $2p-2$ and thus we need an even number of
equations to determine the unknown parameters. So we have two options:

\begin{itemize}
\item[$-$]  \textbf{Even number of loops as input}: In this case we incorporate an even
number ($2p-2$ ) of terms   from the perturbation series to match with corresponding terms 
from the expansion of $_{\text{ }p}F_{p-2}(a_{1},...a_{p}; b_{1}....b_{p-2}%
; -\sigma z)$.

\item[$-$]  \textbf{Odd number of loops as input}: in this case we take odd number ($2p-1$) of loops 
to build odd number of equations and one equation from the large-order constraint:%
\[
\sum_{i=1}^{p}a_{i}-\sum_{i=1}^{p-2}b_{i}-2=b,
\]
to determine the unknown numerator and denominator parameters.
\end{itemize}
So we list resummation results that involve odd or  even number of perturbtive terms   separately.
\subsubsection{ \textbf{Two-loops Resummation for $\nu$}}

For $p=q+2$, the lowest order Hypergeometric approximant for $\nu^{-1}$ is thus:%

\begin{equation}
2 _{\text{ }2}F_{0}\left(  {a_{1},a_{2};\ ; -\frac{3}{N+8}\varepsilon}\right)
= \frac{2}{\Gamma\left(  a_{1}\right)  \Gamma\left(  a_{2}\right)  }\MeijerG*{1}{2}%
{2}{1}{1-a_{1},1-a_{2}}{0}{ -\frac{3}{N+8}\varepsilon}.
\end{equation}
For this resummation function, one needs to determine the two parameters
$a_{1}$and $a_{2}$ by matching the perturbative expansion of $2_{\text{ }%
2}F_{0}(a_{1},a_{2};$ \ $;-\frac{3}{N+8}\varepsilon)$ with the first two terms
in the perturbation series in Eq.(\ref{eps-pert}). In this case we get:
\begin{align}
-\frac{6a_{1}a_{2}}{N+8}  &  =-\frac{N+2}{N+8}\nonumber\\
\frac{9a_{1}(a_{1}+1)a_{2}(a_{2}+1)}{(N+8)^{2}}  &  =-\frac{(N+2)(13N+44)}%
{2(N+8)^{3}},
\end{align}
from which we obtain the results:
\begin{equation}
a_{1}=\frac{-N^{2}-\sqrt{N^{4}+60N^{3}+1636N^{2}+10464N+20032}-42N-152}%
{12(N+8)}%
\end{equation}%
\begin{equation}
a_{2}=\frac{1}{12(N+8)}\left(
\begin{array}
[c]{c}%
\frac{N^{3}}{N+8}+\frac{50N^{2}}{N+8}-2N^{2}+\frac{\sqrt{N^{4}+60N^{3}%
+1636N^{2}+10464N+20032}N}{N+8}\\
+\frac{8\sqrt{N^{4}+60N^{3}+1636N^{2}+10464N+20032}}{N+8}+\frac{488N}%
{N+8}-84N+\frac{1216}{N+8}-304
\end{array}
\right)  .
\end{equation}
To test the accuracy of this two -loops resummation function, let us note that
for $N=1$, the recent Monte Carlo calculation \cite{MC10} gives $\upsilon
=0.63002(10)$. Our two-loops Hypergeometric-Meijer resummation gives the
result $\upsilon=0.66209.$ This result is very reasonable in taking into account that the algorithm is fed with only the first two orders
from the perturbation series as input. For $N=0,$ the a recent accurate
prediction is listed in Ref. \cite{nuN0} as $\nu=0.5875970( 4)$ while our two
loops resummation gives $ \nu=0.60890$. For $N=2$, Monte Carlo
calculations gives $\upsilon=0.6690$ \cite{MC10} while the two-loops gives
$\nu=0.711526$. So it seems that the simple Hypergeometric-Meijer resummation
algorithm we follow in this work gives reasonable results even with very low
orders of perturbation series as   input. It is expected that the resummation of higher orders will improve the accuracy of the results which we will do in the following subsections. 

\subsubsection{\textbf{Three-loops resummation for $\nu$}}

For more accurate results, one can go to the higher three-loops order of
Hypergeometric-Meijer approximants $_{\text{ }3}F_{1}(a_{1},a_{2},a_{3}; b_{1}$
\ $; -\frac{3}{N+8}\varepsilon)$. Although it is parametrized by four
parameters $(a_{1},a_{2},a_{3}$ and $b_{1})$, the use of the large order
constraint \cite{Abo-large}:
\[
\sum_{i=1}^{p}a_{i}-\sum_{i=1}^{p-2}b_{i}-2=b,
\]
leads to the need of three terms only from perturbation series to determine
the parameters. So to determine them $(a_{1},a_{2},a_{3}$ and
$b_{1})$, we solve the set of equations:%

\begin{align}
c_{1}  &  =\frac{2a_{1}a_{2}a_{3}}{b_{1}}\sigma\nonumber\\
c_{2}  &  =\frac{a_{1}(a_{1}+1)a_{2}(a_{2}+1)a_{3}(a_{3}+1)}{b_{1}(b_{1}%
+1)}\sigma^{2}\nonumber\\
c_{3}  &  =\frac{a_{1}(a_{1}+1)(a_{1}+2)a_{2}(a_{2}+1)(a_{2}+2)a_{3}%
(a_{3}+1)(a_{3}+2)}{3b_{1}(b_{1}+1)(b_{1}+2)}\sigma^{3}\\
b  &  =a_{1}+a_{2}+a_{3}-b_{1}-2.\nonumber
\end{align}
The predictions of this order are given in table-\ref{nu3} for different $N$ values and compared to two, four and five
loops resummation results and to the Janke-Kleinert resummation (up to five-loops) in
Ref.\cite{Kleinert-Borel} and the Borel-with conformal mapping  in
Refs.\cite{zin-exp,ON17}. One can easily realize that the convergence has been
greatly improved when moved from two-loops to the three-loops resummation.

\begin{table}[pth]
\caption{{\protect\scriptsize {The two, three, four and five-loops
($\varepsilon-$expansion) Hypergeometric-Meijer resummation for the critical
exponent $\nu$ for the $O(N)$ model compared to the   $\varepsilon^5 $ Janke-Kleinert (JK) resummation results 
(sixth column)  from Ref.\protect\cite{Kleinert-Borel} and the Borel with conformal mapping (BCM) resummation
(seventh column) from Ref.\protect\cite{zin-exp} (first row) and recent results from       Ref.\protect\cite{ON17} (second row).}}}%
\label{nu3}
\begin{tabular}{|l|l|l|l|l|l|l|}
\hline
\multirow{2}{*}{\ \  N \ } & \multicolumn{4}{c|}{This Work} & \ \ \ \ JK$^{ \textsc{\protect\cite{Kleinert-Borel}}}$ & \ \ \ BCM$^{ \textsc{\protect\cite{zin-exp},\cite{ON17}}}$ \\ \cline{2-7} 
 & $_{\text{ }2}F_{0}$:$\ \varepsilon^2$ & $_{\text{ }3}F_{1}$:\ $\varepsilon^3$ & $_{\text{ }3}F_{1}$:\ $\varepsilon^4$ & $_{\text{ }4}F_{2}$:\ $\varepsilon^5$ & \ \ \ \ \ $\varepsilon^5$ &$\ \ \ \ \ \varepsilon^5$ \\ \hline
\ \ 0\ & 0.60890 & 0.58609 & 0.58705 & 0.58714 & 0.5865(13) & \begin{tabular}[c]{@{}l@{}}$0.5875\pm0.0018$\\ 0.5873(13)\end{tabular} \\ \hline
\ \ 1\ & 0.66209 & 0.62502 & 0.62699 & 0.62818 & 0.6268(22) & \begin{tabular}[c]{@{}l@{}}$0.6293 \pm0.0026$\\ 0.6290(20)\end{tabular} \\ \hline
\ \ 2 \ & 0.71153 & 0.66062 & 0.66103 & 0.667225 & 0.6642(111) & \begin{tabular}[c]{@{}l@{}}$0.6685\pm0.0040$\\0.6687(13)\end{tabular} \\ \hline
\ \ 3 \ & 0.75615 & 0.69282 & 0.69303 & 0.70364 & 0.6987(51) & \begin{tabular}[c]{@{}l@{}}$0.7050\pm0.0055$\\ 0.7056(16)\end{tabular} \\ \hline
\ \ 4\ & 0.79557 & 0.72175 & 0.72176 & 0.73692 &\ \ \ \ \ \textbf{\text{\----}}  & \begin{tabular}[c]{@{}l@{}}\ $0.737\pm 0.008$\\ 0.7389(24)\end{tabular} \\ \hline
\end{tabular}
\end{table}
The obvious acceleration of the convergence of the algorithm from
two to three loops is strongly recommending the Hypergeometric-Meijer
resummation algorithm to take a place  among  the preferred   algorithms to resum
divergent series with large order behavior of the form in Eq.(\ref{LOB}). Other features that recommend it for resummation of divergent series is that it does not include any free parameters and of closed form as well.

\subsubsection{\textbf{Four-loops Resummation for $\nu$}}

The Hypergeometric approximants $_{\text{ }3}F_{1}(a_{1},a_{2},a_{3}; b_{1}$
\ $; -\frac{3}{N+8}\varepsilon)$ can also be used to resum the perturbation
series up to four loops but in this case we have to solve the set of equations:
\begin{align}
c_{1}  &  =\frac{2a_{1}a_{2}a_{3}}{b_{1}}\sigma\nonumber\\
c_{2}  &  =\frac{a_{1}(a_{1}+1)a_{2}(a_{2}+1)a_{3}(a_{3}+1)}{b_{1}(b_{1}%
+1)}\sigma^{2}\nonumber\\
c_{3}  &  =\frac{a_{1}(a_{1}+1)(a_{1}+2)a_{2}(a_{2}+1)(a_{2}+2)a_{3}%
(a_{3}+1)(a_{3}+2)}{3b_{1}(b_{1}+1)(b_{1}+2)}\sigma^{3}\\
c_{4}  &  =\frac{a_1(a_1+1)(a_1+2)(a_1+3)a_2(a_2+1)(a_2+2)(a_2+3)a_3(a_3+1)(a_3+2)(a_3+3)}%
{12b_1(b_1+1)(b_1+2)(b_1+3)}\sigma^{4}.\nonumber
\end{align}
The prediction of this order of resummation is also listed in table-\ref{nu3}
where it shows that the accuracy is improving in a systematic way when moving
to higher orders.

\subsubsection{\textbf{Five-loops resummation for $\nu$}}

In this case we use the approximants $_{\text{ }4}F_{2}(a_{1},...,a_{4}%
; b_{1}...b_{4}$ \ $;-\frac{3}{N+8}\varepsilon)$ where the unknown parameters
are determined from the set of equations:
\begin{align}
c_{1}  &  =\frac{2a_{1}a_{2}a_{3}a_{4}\sigma}{b_{1}b_{2}}\nonumber\\
c_{2}  &  =\frac{2a_{1}\left(  a_{1}+1\right)  a_{2}\left(  a_{2}+1\right)
a_{3}a_{4}\left(  a_{3}a_{4}+1\right)  \sigma^{2}}{b_{1}\left(  b_{1}%
+1\right)  b_{2}\left(  b_{2}+1\right)  }\nonumber\\
c_{3}  &  =\frac{a_{1}\left(  a_{1}+1\right)  \left(  a_{1}+2\right)
a_{2}\left(  a_{2}+1\right)  \left(  a_{2}+2\right)  a_{3}a_{4}\left(
a_{3}a_{4}+1\right)  \left(  a_{3}a_{4}+2\right)  \sigma^{3}}{3b_{1}\left(
b_{1}+1\right)  \left(  b_{1}+2\right)  b_{2}\left(  b_{2}+1\right)  \left(
b_{2}+2\right)  }\nonumber\\
c_{4}  &  =\frac{a_{1}\left(  a_{1}+1\right)  \left(  a_{1}+2\right)  \left(
a_{1}+3\right)  \ ......a_{4}\left(  a_{4}+1\right)  \left(  a_{4}+2\right)
\left(  a_{4}+3\right)  \sigma^{4}}{12b_{1}\left(  b_{1}+1\right)  \left(
b_{1}+2\right)  \left(  b_{1}+3\right)  b_{2}\left(  b_{2}+1\right)  \left(
b_{2}+2\right)  \left(  b_{2}+3\right)  },\nonumber\\
c_{5}  &  =\frac{a_{1}\left(  a_{1}+1\right)  \left(  a_{1}+2\right)  \left(
a_{1}+3\right)  \left(  a_{1}+4\right)  ......a_{4}\left(  a_{4}+1\right)
\left(  a_{4}+2\right)  \left(  a_{4}+3\right)  \left(  a_{4}+4\right)
\sigma^{5}}{60b_{1}\left(  b_{1}+1\right)  \left(  b_{1}+2\right)  \left(
b_{1}+3\right)  \left(  b_{1}+4\right)  b_{2}\left(  b_{2}+1\right)  \left(
b_{2}+2\right)  \left(  b_{2}+3\right)  \left(  b_{2}+4\right)  }\\
b  &  =a_{1}+a_{2}+a_{3}+a_{4}-b_{1}-b_{2}-2.\nonumber
\end{align}
For this order, we get even more precise results for the $\nu$-exponent which are also
presented in table-\ref{nu3} and compared to the five-loops resummation from
other algorithms in Refs.\cite{Kleinert-Borel, zin-exp}. Also to compare with other
recent theoretical predictions, for $N=0$, we get the result $\nu=0.587142$
compared to the recent accurate Monte Carlo simulation prediction from Ref. \cite{nuN0} as
$\nu=0.5875970( 4)$. For $N=1$ our five-loops result gives $\nu=0.62818 $ that
can be compared to Monte Carlo calculation that gives $\upsilon=0.63002(10)$
\cite{MC10}. The $N=2$ five-loops resummation in this work gives
$\nu=0.667225$ which is competitive to Monte Carlo calculations of
$\upsilon=0.6690$ in Ref.\cite{MC10}. Also, for $N=3$, our five-loops
resummation gives $\nu=0.703644$ while the recent Monte Carlo prediction gives
$\nu=0.7116(10)$ \cite{MC11}. These results show clearly that our five-loops resummation
results are competitive   either to five-loops resummation from other algorithms or to recent numerical methods.

To get an impression about the stability of the algorithm predictions for
higher $N$ values, we list in table-\ref{tab:5lresN} our five-loops
resummation ($_{\text{ }4}F_{2}(a_{1},a_{2},a_{3},a_{4};b_{1},b_{2}\ ;-\sigma
z)$) results  for $N=6,8,10,12$ and compared them to other theoretical predictions.

\begin{table}[pth]
\caption{{\protect\scriptsize {The 5-loops Hypergeometric-Meijer resummation
($_{4}F_{2}$ approximant) of the critical exponent $\nu$ for the $O(N)$ model for
$N=6,8,10$ and $12$ compared to  other theoretical predictions. Ref.\protect\cite{Eta42}
used the strong coupling resummation and  Ref.\protect\cite{Bstrab} is a conformal bootstrap calculation where
we used $\Delta_{s}=2-3/\nu$ to get the listed results. In Ref.\protect\cite{ON12},
numerical calculations are used to predict the critical exponents and in
Ref.\protect\cite{OTD} the the optimally truncated direct summation of  pseudo-$\epsilon$ expansion  ($\tau$,OTDS) has been used where
we obtained the listed result via the relation $\alpha=2-D \nu$.}}}%
\label{tab:5lresN}
 \begin{tabular}{|l|l|l|l|l|}
\hline
\multicolumn{1}{|c|}{N} & \ \ \ \  \ \ \ 6 &\ \ \ \  \ \ \ 8 & \ \ \ \  \ \ \ 10 & \ \ \ \  \ \ \ 12 \\ \hline
\begin{tabular}[c]{@{}l@{}} This work\\ \ \ $_{\text{ }4}F_{2}$ :\ $\varepsilon^5$\end{tabular} &\ \ \ \  0.79331  & \ \ \ \ 0.83692 &\ \ \ \  0.88809 & \ \ \ \ 0.89472\\ \hline
\begin{tabular}[c]{@{}l@{}}\ \ Other\\ calculations\end{tabular} & \begin{tabular}[c]{@{}l@{}}$\ \ \ \ 0.790^{\textsc{\protect\cite{Eta42}}}$\\ $0.78431^{+0.032}_{-0.033} \ ^{\textsc{\protect\cite{Bstrab}}}$\end{tabular} & \begin{tabular}[c]{@{}l@{}}$\ \ \ \ 0.829^{\textsc{\protect\cite{Eta42}}}$\\ \ \   $0.8183 ^{\textsc{\protect\cite{OTD}}}$\end{tabular} & \begin{tabular}[c]{@{}l@{}}$\ \ \ \ 0.866^{\textsc{\protect\cite{Eta42}}}$\\ $0.88417^{+0.000}_{-0.0008} \ \ ^{\textsc{\protect\cite{Bstrab}}}$\end{tabular} & \begin{tabular}[c]{@{}l@{}}$\ \ \ \ 0.890^{\textsc{\protect\cite{Eta42}}}$\\ \ \ \ \  $0.93279\ ^{\textsc{\protect\cite{ON12}}}$\end{tabular} \\ \hline
\end{tabular}%

\end{table}
\subsection{ Resummation of Four and Five-loops series for $\eta$ exponent}

For the critical exponent $\eta$ of the $O(N)$ model, the $\varepsilon
$-expansion up to five loops is given by \cite{Kleinert-Borel}
\begin{equation}
\eta =\varepsilon^{2}\left(  d_{2}+d_{3}\varepsilon+d_{4}\varepsilon^{2}%
+d_{5}\varepsilon^{3}\right)  +O(\varepsilon^{6}) \label{eta-pert}%
\end{equation}
where%

\begin{align}
d_{2} &  =\frac{(N+2)\ }{2(N+8)^{2}}\nonumber\\
d_{3} &  =\frac{(N+2)\left(  -N^{2}+56N+272\right)  }{8(N+8)^{4}}\nonumber\\
d_{4} &  =\frac{(N+2)}{32(N+8)^{6}}\{-5N^{4}-230N^{3}+1124N^{2}%
-384(N+8)(5N+22)\zeta(3)+17920N+46144\}\nonumber\\
d_{5} &  =-\frac{(N+2)}{128(N+8)^{8}}\{13N^{6}+946N^{5}+27620N^{4}%
+121472N^{3}-262528N^{2}-2912768N\nonumber\\
&  -5120(N+8)^{2}\left(  2N^{2}+55N+186\right)  \zeta(5)\frac{64}{5}\pi
^{4}(N+8)^{3}(5N+22)-5655552\nonumber\\
&  -16(N+8)\left(  N^{5}+10N^{4}+1220N^{3}-1136N^{2}-68672N-171264\right)
\zeta(3)-5655552\}
\end{align}
and the large-order for $\eta$ of this model takes the form in Eq.(\ref{LOB})
where \cite{Kleinert-Borel}
\[
\sigma=\frac{3}{N+8}\text{ and \ }b=3+\frac{N}{2}.
\]
Note that the factored series $\left(  d_{2}+d_{3}\varepsilon+d_{4}%
\varepsilon^{2}+d_{5}\varepsilon^{3}\right)  +O(\varepsilon^{6})$ has the
large-order parameters \cite{Kleinert-Borel}
\[
\sigma=\frac{3}{N+8}\text{ and \ }b=5+\frac{N}{2}.
\]
The lowest order approximant is thus $_{2}F_{0}$ which in this case is a
four-loops approximant. 

\subsubsection{\textbf{Four-loops resummation for $\eta$}}

The Hypergeometric-Meijer approximant is then:
\begin{align}
\eta &  =d_{2}(N)\varepsilon^{2}\,_{2}F_{0}(a_{1},a_{2};  ;-\sigma\varepsilon
)\nonumber\\
&  =\frac{d_{2}(N)\varepsilon^{2}}{\Gamma\left(  a_{1}\right)  \Gamma\left(
a_{2}\right)  }\MeijerG*{1}{2}{2}{1}{1-a_{1},1-a_{2}}{0}{ -\frac{3}{N+8}\varepsilon}%
\end{align}
The resummation results of that order are shown in table-\ref{eta45}. The results are reasonable but since  the Hypergeometric approximant  $_{2}F_{0}$ has few number of parameters, it is expected that the  improvement of  the results needs higher loops to be incorporated. 
\subsubsection{\textbf{The $\eta$ five-loop resummation}}

In this case the Hypergeometric approximant is
\begin{equation}
\eta=d_{2}(N)\varepsilon^{2}\,_{3}F_{1}(a_{1},a_{2},a_{3};b_{1};-\sigma
\varepsilon). 
\end{equation}
To determine the four unknown parameters we use the equations:%
\begin{align}
d_{3}  &  =d_{2}\frac{a_{1}a_{2}a_{3}}{b_{1}}\sigma\nonumber\\
d_{4}  &  =d_{2}\frac{a_{1}\left(  1+a_{1}\right)  a_{2}\left(  1+a_{2}%
\right)  a_{3}\left(  1+a_{3}\right)  }{b_{1}\left(  1+b_{1}\right)  }%
\sigma\nonumber\\
d_{5}  &  =d_{2}\frac{a_{1}\left(  1+a_{1}\right)  \left(  2+a_{1}\right)
a_{2}\left(  1+a_{2}\right)  \left(  2+a_{2}\right)  a_{3}\left(
1+a_{3}\right)  \left(  2+a_{3}\right)  }{b_{1}\left(  1+b_{1}\right)  \left(
2+b_{1}\right)  }\sigma\\
5 & +\frac{N}{2}   =a_{1}+a_{2}+a_{3}-b_{1}-2.\nonumber
\end{align}
Accordingly, the Hypergeometric-Meijer approximant for this order is given by:%
\begin{align}
\eta &  =d_{2}(N)\varepsilon^{2}\,_{3}F_{1}({a_{1}},{a_{2}},{a_{3}};{b_{1}%
};-\sigma\varepsilon)\nonumber\\
&  =d_{2}(N)\varepsilon^{2}\frac{\Gamma(b_{1})}{\Gamma\left(  a_{1}\right)
\Gamma\left(  a_{2}\right)  \Gamma\left(  a_{3}\right)  }\MeijerG*{1}{3}{3}{2}%
{1-a_{1},1-a_{2},1-a_{3}}{0,1-b_{1}}{-\frac{3}{N+8}\varepsilon}%
\end{align}
Our predictions that incorporate   the fourth and fifth orders of divergent series of the $\eta$-exponent 
are listed in table-\ref{eta45} . It is very clear that
the simple algorithm we follow gives  accurate results for few terms from
the perturbation series as input. This can be more elaborated by looking at
the large number of estimates for critical exponents in Ref.\cite{pelsito} too. In fact, for the same order of perturbation series involved,  the precision of resummation results for $\eta$ are always less than that in $\nu$ or $\omega$ because the lowest order in the perturbation series of $\eta$ is $\varepsilon^2$ and thus always approximated by Hypergeometric approximants of fewer parameters than that for $\nu$ or $\omega$.

\begin{table}[pth]
\caption{{\protect\scriptsize {The four and five-loops ($\varepsilon
-$expansion) Hypergeometric-Meijer resummation for the critical exponent
$\eta$ for the $O(N)$ model. We compared the results to Janke-Kleinert Resummation  for five-loops $\varepsilon$-expansion in Ref.\protect\cite{Kleinert-Borel} and the Borel with conformal mapping resummation  from
Ref.\protect\cite{zin-exp} (first) and Ref.\protect\cite{ON17} (second)} }}

\label{eta45}
\begin{tabular}{|l|l|l|l|l|}
\hline
\multicolumn{1}{|c|}{\multirow{2}{*}{\ \ \ N\ \ }} & \multicolumn{2}{l|}{\ \ \ This work} & \ \ \ JK$^\textsc{{\protect\cite{Kleinert-Borel}}}$ & \ \ \ BCM $^\textsc{{\protect\cite{zin-exp},\cite{ON17}}}$\\ \cline{2-5} 
\multicolumn{1}{|c|}{} & $_{\text{ }2}F_{0}$ :\ $\varepsilon^4$ & $_{\text{ }3}F_{1}$ :\ $\varepsilon^5$ & $\ \ \ \ \ \varepsilon^5$ & \ \ \ \ \ \ \ $\varepsilon^5$\\ \hline
\ \ \ 0\ \ & 0.02804 & 0.03111 & 0.0344(42) & \begin{tabular}[c]{@{}l@{}}$0.0300 \pm0.0060$\\ 0.0314(11)\end{tabular} \\ \hline
\ \ \ 1\ \ & 0.03286 & 0.03615 & 0.0395(43) & \begin{tabular}[c]{@{}l@{}}$0.0360\pm0.0060$\\ 0.0366(11)\end{tabular} \\ \hline
\ \ \ 2\ \ & 0.03475 & 0.03791 & 0.0412(41) & \begin{tabular}[c]{@{}l@{}}$0.0385 \pm0.0065$\\ 0.0384(10)\end{tabular} \\ \hline
\ \ \ 3\ \ & 0.03498 & 0.03781 & 0.0366(20) & \begin{tabular}[c]{@{}l@{}}$0.0380 \pm0.0060$\\ 0.0382(10)\end{tabular} \\ \hline
\ \ \ 4\ \ & 0.034274 & 0.03668 & \ \ \ \textbf{------} & \begin{tabular}[c]{@{}l@{}}$0.036 \pm 0.004$\\ 0.0370(9)\end{tabular} \\ \hline
\end{tabular}%

\end{table}
\subsection{Resummation of the exponent $\omega$}

For the exponent $\omega$ we have the five-loops perturbation series as:
\begin{equation}
\omega=\varepsilon+e_{2}\varepsilon^{2}+e_{3}\varepsilon^{3}+e_{4}%
\varepsilon^{4}+e_{5}\varepsilon^{5}+O(\varepsilon^{6}), \label{omega-pert}%
\end{equation}
where \cite{zin-borel}%

\begin{align}
e_{2}& =-\frac{3(3N+14)}{(N+8)^{2}},  \nonumber \\
e_{3}& =\frac{(33N^{3}+538N^{2}+4288N+9568+\zeta \lbrack 3](N+8)96(5N+22))}{%
4(N+8)^{4}},  \nonumber \\
e_{4}& =\frac{1}{16(N+8)^{6}}\{5N^{5}-1488N^{4}-46616N^{3}-1920(N+8)^{2}%
\left( 2N^{2}+55N+186\right) \zeta (5)  \nonumber \\
& 
\begin{array}{c}
\text{ \ \ \ \ \ \ \ \ \ \ \ \ \ \ \ \ \ \ \ }-419528N^{2}-96(N+8)\left(
63N^{3}+548N^{2}+1916N+3872\right) \zeta (3) \\ 
\text{ \ \ \ \ \ \ \ }-1750080N+\frac{16}{5}\pi
^{4}(N+8)^{3}(5N+22)-2599552\},%
\end{array}
\nonumber \\
e_{5}& =\frac{1}{64(N+8)^{8}}%
\{13N^{7}+7196N^{6}+240328N^{5}+3760776N^{4}+38877056N^{3}  \nonumber \\
& +112896(N+8)^{3}\left( 14N^{2}+189N+526\right) \zeta (7)+223778048N^{2} 
\nonumber \\
& +660389888N+752420864-\frac{640}{63}\pi ^{6}(N+8)^{4}\left(
2N^{2}+55N+186\right)   \nonumber \\
& -\frac{16}{5}\pi ^{4}(N+8)^{3}\left( 63N^{3}+548N^{2}+1916N+3872\right)  
\nonumber \\
& +256(N+8)^{2}\zeta (5)\left(
305N^{4}+7386N^{3}+45654N^{2}+143212N+226992\right)   \nonumber \\
& -768(N+8)^{2}\left( 6N^{4}+107N^{3}+1826N^{2}+9008N+8736\right) \zeta
(3)^{2}  \nonumber \\
& -16(N+8)\zeta (3)[9N^{6}-1104N^{5}-11648N^{4}-243864N^{3}-2413248N^{2} 
\nonumber \\
& -9603328N-14734080]\}
\end{align}

and the large-order parameters for that exponent are
\[
\sigma=\frac{-3}{N+8}\text{ and \ }b=5+\frac{N}{2}.
\]
The two-loops resummation gives reasonable but not precise results so in the
  following, we shall list the resummation of three, four and five loops.

\subsubsection{\textbf{ Three-loops Resummation for $\omega$}}

The three-loops Hypergeometric approximant is:
\begin{equation}
\omega\approx\,_{3}F_{1}(a_{1},a_{2},a_{3};b_{1};-\sigma\varepsilon)-1, 
\end{equation}
where
\begin{align}
1  &  =\frac{a_{1}a_{2}a_{3}\ \sigma}{b_{1}b_{2}},\nonumber\\
e_{2}  &  =\frac{a_{1}\left(  a_{1}+1\right)  a_{2}\left(  a_{2}+1\right)
a_{3}\left(  a_{4}+1\right)  \sigma^{2}}{2b_{1}\left(  b_{1}+1\right)
b_{2}\left(  b_{2}+1\right)  },\nonumber\\
e_{3}  &  =\frac{a_{1}\left(  a_{1}+1\right)  \left(  a_{1}+2\right)
a_{2}\left(  a_{2}+1\right)  \left(  a_{2}+2\right)  a_{3}\ \left(
a_{3}\ +1\right)  \left(  a_{3}\ +2\right)  \sigma^{3}}{6b_{1}\left(
b_{1}+1\right)  \left(  b_{1}+2\right)  },\nonumber\\
b  &  =a_{1}+a_{2}+a_{3}\ -b_{1}-b_{2}-2. \label{om3L}%
\end{align}
The solutions of these equations are then substituted  in the following Meijer-G function :

\begin{equation}
\omega \approx\frac{\Gamma(b_{1})}{\Gamma\left(  a_{1}\right)  \Gamma\left(
a_{2}\right)  \Gamma\left(  a_{3}\right)  } \MeijerG*{1}{3}{3}{2}{1-a_{1}%
,1-a_{2},1-a_{3}}{0,1-b_{1}}{-\frac{3}{N+8}\varepsilon}-1%
\end{equation}

\subsubsection{  \textbf{The $\omega$ four-loops Resummation } }

In this case also we use the approximant $_{3}F_{1}(a_{1},a_{2},a_{3}%
;b_{1};-\sigma\varepsilon)$ but   we replace the fourth equation in the
set in Eqs.(\ref{om3L}) by:%
\begin{equation}
e_{4}=\frac{a(a+1)(a+2)(a+3)b(b+1)(b+2)(b+3)c(c+1)(c+2)(c+3)}%
{12d(d+1)(d+2)(d+3)}\sigma^{4}%
\end{equation}

\subsubsection{\textbf{$\omega$ five-loops approximant}}

The Hypergeometric function that can accommodate five-loops is $_{4}%
F_{2}(a_{1},a_{2},a_{3},a_{4};b_{1},b_{2};-\sigma\varepsilon)$ where we use the
constraint on the large order parameters:%

\[
b=a_{1}+a_{2}+a_{3}+\ a_{4}-b_{1}-b_{2}-2.
\]
Accordingly, the fifth order resummation for $\omega$ is
\begin{equation}
\omega \approx\left(  \frac{\Gamma(b_{1})\Gamma(b_{2})}{\Gamma\left(  a_{1}\right)
\Gamma\left(  a_{2}\right)  \Gamma\left(  a_{3}\right)  \Gamma\left(
a_{4}\right)  } \MeijerG*{1}{4}{4}{3}{1-a_{1},1-a_{2},1-a_{3},1-a_{4}}{0,1-b_{1}%
,1-b_{2}}{-\frac{3}{N+8}\varepsilon}-1\right )   
\end{equation}
In table-\ref{omega345}, we compared our results to predictions  from the
Janke-Kleinert  Resummation for five-loops $\varepsilon$-expansion in
Ref.\cite{Kleinert-Borel} and Borel with conformal mapping   in Refs.\cite{zin-exp,ON17} for $N=0,1,2,3$ and $4$. Again, the
comparison shows that the algorithm we follow gives very accurate results from
few orders of the perturbation series as input.

\begin{table}[pth]
\caption{{\protect\scriptsize{The three, four and five-loops Hypergeometric-Meijer resummation for
the critical exponent $\omega$ compared to five-loops resummation from
Ref.\protect\cite{Kleinert-Borel} (fifth column) and the Borel with conformal mapping
resummation (sixth column) from Refs.\protect\cite{zin-exp,ON17}.  }}}
\label{omega345}
\begin{tabular}
[c]{|l|l|l|l|l|l|}\hline
\ \ N\ \  & \multicolumn{1}{c|}{%
\begin{tabular}
[c]{@{}c@{}}%
$\  _{\text{ }3}F_{1}$\\
This work: $\varepsilon^3$
\end{tabular}
} &
\begin{tabular}
[c]{@{}l@{}}%
$\ \ _{\text{ }3}F_{1}$\\
This work: $\varepsilon^4$
\end{tabular}
&
\begin{tabular}
[c]{@{}l@{}}%
$\ \ _{\text{ }4}F_{2}$\\
This work: $\varepsilon^5$
\end{tabular}
&
\begin{tabular}
[c]{@{}l@{}}%
\ JK$^\textsc{{\protect\cite{Kleinert-Borel}}}$:\  $\varepsilon^5$
\end{tabular}
&
\begin{tabular}
[c]{@{}l@{}}%
\  BCM$^\textsc{{\protect\cite{zin-exp},\protect\cite{ON17}}}$:\  $\varepsilon^5$
\end{tabular}
\\\hline
\ \ 0\ \  & 0.86128 & 0.80054 & 0.85086 & 0.817(21) &\begin{tabular}[c]{@{}l@{}}$0.828 \pm 0.023 $\\ 0.835(11)\end{tabular}   \\\hline
\multicolumn{1}{|c|}{1} & 0.85628 & 0.79559 & 0.83178 & 0.806(13)  & \begin{tabular}[c]{@{}l@{}}$0.814 \pm0.018$\\0.818(8)\end{tabular} 
 \\\hline
\ \ 2\ \  & 0.85233 & 0.79290 & 0.81329 & 0.800(13) & \begin{tabular}[c]{@{}l@{}}$0.802 \pm0.018$\\ 0.803(6) \end{tabular} \\ \hline
\ \ 3\ \  & 0.84979 & 0.79258 & 0.79928 & 0.796(11)  & \begin{tabular}[c]{@{}l@{}} $0.794 \pm0.018$\\ 0.797(7) \end{tabular} \\ \hline
\ \ 4\ \  & 0.910678 & 0.79416 & 0.79249 &\ \ \ \ \ \textbf{\text{\----}} &\begin{tabular}[c]{@{}l@{}} $0.795 \pm 0.030$\\ 0.795(6) \end{tabular} \\ \hline
\end{tabular}
\end{table}

\subsection{Resummation of the $\varepsilon-$expansion for the critical
coupling}

In the way to get the $\varepsilon$-expansion for the critical exponents one
has to obtain the dependance of the critical coupling on $\varepsilon$ first. The
expansion for the critical coupling $g_{c}$ up to fifth order is given by
\cite{Kleinert-Borel}:%

\begin{align}
\text{For }N  &  =0\Rightarrow g_{c}\left(  \varepsilon\right)
\approx 0.375\varepsilon+0.246\varepsilon^{2}-0.180\varepsilon^{3}+0.368\varepsilon
^{4}-1.258\varepsilon^{5},\nonumber\\
\text{For }N  &  =1\Rightarrow g_{c}\left(  \varepsilon\right)
\approx 0.333\varepsilon+0.210\varepsilon^{2}-0.138\varepsilon^{3}+0.269\varepsilon
^{4}-0.8445\varepsilon^{5},\nonumber\\
\text{For }N  &  =2\Rightarrow g_{c}\left(  \varepsilon\right)
\approx 0.3\varepsilon+0.18\varepsilon^{2}-0.108\varepsilon^{3}+0.205\varepsilon
^{4}-0.591\varepsilon^{5},\nonumber\\
\text{For }N  &  =3\Rightarrow g_{c}\left(  \varepsilon\right)
 \approx0.273\varepsilon+0.156\varepsilon^{2}-0.086\varepsilon^{3}+0.162\ \varepsilon
^{4}-0.430\varepsilon^{5},\\
\text{For }N  &  =4\Rightarrow g_{c}\left(  \varepsilon\right)
\approx \frac{1}{4}\varepsilon+\frac{13}{96}\varepsilon^{2}-0.0707\varepsilon^{3}+0.130\ \varepsilon
^{4}-0.322\varepsilon^{5}\nonumber%
\end{align}
while the large order parameters are $\sigma=\frac{3}{N+8}\text{ and
\ }b=4+\frac{N}{2}$. The third order approximation takes the form $\,_{3}%
F_{1}(a_{1},a_{2},a_{3};b_{1};-\sigma\varepsilon)-1$ while the fourth order
takes the same form except in the equations determining the parameters we use
the large order constraint $a_{1}+a_{2}+a_{3}-b_{1}-2=b$. For the five-loops
resummation we resummed the series
\begin{equation}
\frac{g_{c}\left(  \varepsilon\right)  }{\varepsilon}=f_{1}+f_{2}%
\varepsilon+f_{3}\varepsilon^{2}+f_{4}\varepsilon^{3}+f_{5}\varepsilon^{4},
\end{equation}
for $N=1,2,3$ and $4$ using the Hypergeometric approximant $f_{1}$
$_{3}F_{1}(a_{1},a_{2},a_{3};b_{1};\sigma\varepsilon)$. For $N=0$, however, we
resummed the subtracted series $\frac{g_{c}\left(  \varepsilon\right)
-f_{1}\varepsilon}{f_{2}\varepsilon^{2}}=1+f_{3}\varepsilon+f_{4}%
\varepsilon^{2}+f_{5}\varepsilon^{3}$ using the Hypergeometric approximant:
\begin{equation}
g_{c}\left(  \varepsilon\right)  =f_{1}\varepsilon+f_{2}\varepsilon^{2}\text{
}_{3}F_{1}(a_{1},a_{2},a_{3};b_{1};\sigma\varepsilon),
\end{equation}
with the constraint $a_{1}+a_{2}+a_{3}-b_{1}-2=b+2.$ Such technical steps are
well known in resummation techniques \cite{Kleinert-Borel, Prd-GF} which can be
used in case no solution has been found for the equations defining the
parameters. The prediction of these orders are shown in table-\ref{gc345} and
compared with other resummation results from Refs.\cite{zin-exp,Kleinert-Borel,Eta42,Eta4}.

\begin{table}[pth]
\caption{{\protect\scriptsize { The three, four and five-loops
Hypergeometric-Meijer resummation of the critical coupling $g_{c}$ for the
$O(N)$-model with $N=0,1,2,3$ and $4$. The result from Ref.\protect\cite{zin-exp} in
the last column (scaled by a factor $\frac{3}{N+8}$ because of different
normalizations) and $SC$ refers to strong coupling resummation algorithm. }}}%
\label{gc345}%
\begin{tabular}
[c]{|l|l|l|l|l|l|}\hline
\ \ N\ \  & \multicolumn{1}{c|}{%
\begin{tabular}
[c]{@{}c@{}}%
$_{\text{ }3}F_{1}$\\
This work: $\varepsilon^3$
\end{tabular}
} &
\begin{tabular}
[c]{@{}l@{}}%
$_{\text{ }3}F_{1}$\\
This work: $\varepsilon^4$
\end{tabular}
&
\begin{tabular}
[c]{@{}l@{}}%
$_{\text{ }3}F_{1}$\\
This work: $\varepsilon^5$
\end{tabular}
&
\begin{tabular}
[c]{@{}l@{}}%
\ \ \ \ \ \ \ JK\textbackslash SC
\end{tabular}
&
\begin{tabular}
[c]{@{}l@{}}%
  \ \ \ \ \ \ \ \ {\scriptsize\protect { BCM$^\textsc{{\protect\cite{zin-exp}}}$}}
\end{tabular}
\\\hline
\ \ 0\ \  & 0.54035  & 0.54684 & 0.49007 & 0.5408(83) 
,{\scriptsize\protect {JK$^\textsc{{\protect\cite{Kleinert-Borel}}}$}} & $0.52988 \pm0.00225$\\\hline
\multicolumn{1}{|c|}{1} & 0.47883  & 0.48475 & 0.48462 &
0.4810(91),{\scriptsize\protect {JK$^\textsc{{\protect\cite{Kleinert-Borel}}}$}} & $0.47033 \pm
0.001$\\\hline
\ \ 2\ \  & 0.42779  & 0.43322 & 0.43429 &
0.5032(239),{\scriptsize\protect {JK$^\textsc{{\protect\cite{Kleinert-Borel}}}$}} & $0.4209 \pm
0.001$\\\hline
\ \ 3\ \  & 0.36955  & 0.39006 & 0.39214 &
0.3895(71),{\scriptsize\protect {JK$^\textsc{{\protect\cite{Kleinert-Borel}}}$}} & $0.37936 \pm
0.001$ \\\hline
\ \ 4\ \  & 0.34921 & 0.35187  & 0.35638 &
0.34375, {\scriptsize\protect {SC$^\textsc{{\protect\cite{Eta42}}}$}} &
$0.34425\pm
0.00125 $\\\hline
\end{tabular}
\end{table}

\section{Six-Loops Hypergeometric-Meijer resummation of the critical exponents $\nu,\eta$
and  $\omega$  }\label{sixL}

In Ref.\cite{ON17}, the six-loops order of  the renormalization group functions has been obtained and resummed using Borel
with conformal mapping algorithm. The work  led to the  improvement of the previous
resummation predictions of the  five-loops order in Refs. \cite{Kleinert-Borel,
zin-exp}.  This six-loops order of perturbation series  represents a good test for the accuracy and stability of our resummation algorithm. We shall thus  extend our work in the previous section to incorporate the six-loops weak-coupling data to compare with the recent results of Borel  resummation and numerical predictions.  
 
\begin{table}[pth]
\caption{{\protect\scriptsize { The six-loops Hypergeometric-Meijer
resummation (first) for the critical exponent $\nu,\eta$ and $\omega$ for $O(N)$-model with $N=0,1,2,3$
and $4$. The results are compared to recent Borel with conformal mapping  (second)  resummation in
Ref.\protect\cite{ON17} and also recent Monte Carlo   simulations methods (third).  }}}%
\label{nu6}
\begin{tabular}{|l|l|l|l|l|}
\hline
\ \ N\ \ & $\ \ \ \ \ \  \nu$ & $\ \ \ \ \ \ \eta$ & $\ \ \ \ \ \omega$ & Reference \\ \hline
\ \ 0\ \ & \begin{tabular}[c]{@{}l@{}}0.58744\\  0.5874(3)\\  0.5875970(4)\end{tabular} & \begin{tabular}[c]{@{}l@{}}0.03034\\  0.0310(7)\\  0.031043(3)\end{tabular} & \begin{tabular}[c]{@{}l@{}}0.85559\\ 0.841(13)\\ 0.904(5)\end{tabular} & \begin{tabular}[c]{@{}l@{}}This work\\ \ \ \ \ \protect\cite{ON17}\\ \ \ \ \ \protect\cite{nuN0}\end{tabular} \\ \hline
\ \ 1\ \ & \begin{tabular}[c]{@{}l@{}}0.62937\\ 0.6292(5)\\  0.63002(10)\end{tabular} & \begin{tabular}[c]{@{}l@{}}0.03545\\ 0.0362(6)\\ 0.03627(10)\end{tabular} & \begin{tabular}[c]{@{}l@{}}0.82929 \\  0.820(7)\\  0.832(6)\end{tabular} & \begin{tabular}[c]{@{}l@{}}This work\\ \ \ \ \ \protect\cite{ON17}\\ \ \ \ \  \protect\cite{MC10}\end{tabular} \\ \hline
\ \ 2\ \ & \begin{tabular}[c]{@{}l@{}}0.66962\\ 0.6690(10)\\  0.6717(1)\end{tabular} & \begin{tabular}[c]{@{}l@{}}0.03733\\  0.0380(6)\\  0.0381(2)\end{tabular} & \begin{tabular}[c]{@{}l@{}}0.80580 \\  0.804(3)\\ 0.785(20)\end{tabular} & \begin{tabular}[c]{@{}l@{}}This work\\ \ \ \ \ \protect\cite{ON17}\\  \ \ \ \ \protect\cite{MCN2}\end{tabular} \\ \hline
\ \ 3\ \ & \begin{tabular}[c]{@{}l@{}}0.70722\\ 0.7059(20)\\  0.7116(10)\end{tabular} & \begin{tabular}[c]{@{}l@{}}0.037301\\ 0.0378(5)\\  0.0378(3)\end{tabular} & \begin{tabular}[c]{@{}l@{}}0.79272 \\ 0.795(7)\\ 0.791(22)\end{tabular} & \begin{tabular}[c]{@{}l@{}}This work\\ \ \ \ \ \protect\cite{ON17}\\  \ \ \ \ \protect\cite{MC11}\end{tabular} \\ \hline
\ \ 4\ \ & \begin{tabular}[c]{@{}l@{}}0.74151\\  0.7397(35)\\ 0.750(2)\end{tabular} & \begin{tabular}[c]{@{}l@{}}0.03621\\ 0.0366(4)\\  0.0360(3)\end{tabular} & \begin{tabular}[c]{@{}l@{}}0.76793 \\  0.794(9)\\ 0.817 (30)\end{tabular} & \begin{tabular}[c]{@{}l@{}}This work\\ \ \ \ \ \protect\cite{ON17}\\ \ \ \ \ \protect\cite{MC11}\end{tabular} \\ \hline
\end{tabular}%

\end{table}
A different $\varepsilon$ has been used in Ref.\cite{ON17}  as the   space-time dimension has been set as $d-2\varepsilon$. Accordingly, the $n^{th}$ coefficients in
each perturbation series has to be divided by $2^{n}$ to keep the definition used in our work ( $d-\varepsilon$). For the critical
exponent $\nu$ we then have
\begin{equation}
\nu^{-1}=2+\sum_{i=1}^{6}c_{i}\varepsilon^{i}+O\left(  \varepsilon^{7}\right)
,
\end{equation}
where the first five coefficients ( $c_{i}$) are given by Eq.(\ref{ccoef}) while the sixth
coefficients are given in table-\ref{coeff6}.
\begin{table}[pth]
\caption{{\protect\scriptsize { The coefficients of the sixth order in the $\varepsilon$-expansion from  
Ref.\protect\cite{ON17} but scaled properly to match with the choice  $d-\varepsilon$ of the space-time dimension in our work while in Ref.\protect\cite{ON17} the choice was  $d-2\varepsilon$. In this table $c_6$ for $\nu^{-1}$, $d_6$ for $\eta$ and $e_6$ for $\omega$ series respectively.}}}%
\label{coeff6}
\begin{tabular}{|l|l|l|l|l|l|}
\hline
\ \ N & \ \ \ 0 & \ \ \ 1 & \ \ \ 2 & \ \ \ 3 & \ \ \ 4 \\ \hline
\ \ $c_6$ & -3.856 & -3.573 & -3.103 & -2.639 & -2.234 \\ \hline
\ \ $d_6$ & -0.0907 & -0.0813  & -0.0686 & -0.0570 & -0.0474 \\ \hline
\ \ $e_6$ & -130.00 &  -93.111 & -68.777 & -52.205 & -40.567 \\ \hline
\end{tabular}
\end{table}
Accordingly we use the approximant $2\ _{4}F_{2}(a_{1},a_{2},a_{3},a_{4}%
;b_{1},b_{2};-\sigma\varepsilon)$ for the resummation of the $\nu^{-1}$ series above. In table \ref{nu6},
one can realize that our six-loop resummation for the critical exponent $\nu$
is very competitive either to the six-loops Borel with conformal mapping algorithm in Ref.\protect\cite{ON17} or Monte Carlo calculations ( ours   are closer to numerical results).

For the critical exponent $\eta$, we have the series up to  fifth order  in Eq.(\ref{eta-pert}) and we add the sixth 
coefficient  from  Ref.\cite{ON17} as shown in table-\ref{coeff6}. The Hypergeometric approximant $\text{ }_{3}F_{1}$ has been used for the resummation of the six-loops perturbation series of $\eta$ and its resummation   results are presented in table \ref{nu6} too.

For the critical exponent $\omega$,   the sixth coefficients $e_6$ are listed in Table-\ref{coeff6}. In this case we use  the approximant $_{4}F_{2}(a_{1},a_{2},a_{3},a_{4};b_{1}%
,b_{2};-\sigma\varepsilon)-1$ which in turn results in the last column  in table \ref{nu6}. Note that when there exist no solution for the
set of equations determining the parameters we resort to successive subtraction of the perturbation series \cite{Kleinert-Borel,Prd-GF}.

\section{Resummation of the the seven-loops  coupling-series for  $\beta$, $\gamma_{m^2}$ and $\gamma_{\phi}$
Renormalization group functions}\label{sevenL}
In the minimal subtraction scheme, Oliver Schnetz has obtained the seven-loops order of the renormalization group functions  $\beta$, $\gamma_{m^2}$ and $\gamma_{\phi}$ for the $O(N)$-symmetric model \cite{7L}. Here $\gamma_{m^2}$ is the mass anomalous dimension while  $\gamma_{\phi}$ represents the field anomalous dimension. In the following we list our resummation results for $N=0,1,2,3$ and $4$ while the results are compared to recent calculations from different techniques in tables \ref{7L0}, \ref{7L1},\ref{7L2}, \ref{7L3} and  \ref{7L4}. Note that for the $g$-series, the large order parameters for the $O(N)$-symmetric model are $\sigma=1$ and $b_{\beta}=3+N/2$  ,  $b_{\omega}=4+N/2$, $b_{\gamma_{\phi}}=2+N/2$  and $b_{\gamma_{m^2}}=3+N/2$  \cite{Kleinert-Borel} where $\omega=\beta^\prime_g $.

\subsection{Resummation results for self-avoiding walks $(N=0)$}
For $N=0$ and in three dimensions, the seven-loops order for the $\beta$-function is given by:
\begin{equation}
\beta\approx-  g+2.667g^{2}- 4.667g^{3}+25.46g^{4}-200.9g^{5}+2004g^{6}- 23315g^{7}+303869g^{8}.
\end{equation}
We resummed this series using the approximant $\left(_{5}F_{3}(a_{1},a_{2},a_{3}%
,a_{4},a_{5};b_{1},b_{2},b_{3};-g)-1\right)$ which resulted in the Meijer-G approximant of the form:
\begin{equation}
\beta=  \frac{\Gamma(b_{1})\Gamma(b_{2})\Gamma(b_{3})}{\Gamma\left(  a_{1}\right)
\Gamma\left(  a_{2}\right)  \Gamma\left(  a_{3}\right)  \Gamma\left(
a_{4}\right)  \Gamma\left(
a_{5}\right)} \MeijerG*{1}{5}{5}{4}{1-a_{1},1-a_{2},1-a_{3},1-a_{4},1-a_{5}}{0,1-b_{1}%
,1-b_{2},1-b_{3}}{-g } -1.
\end{equation}
The critical coupling is obtained from the zero of the $\beta$-function where we found $g_c=0.53430$. The series for  correction to scaling critical exponent $\omega$ is obtained from differentiating the above series with respect to $g$ and it has been resummed using the approximant $\left(-_{5}F_{3}(a_{1},a_{2},a_{3}%
,a_{4},a_{5};b_{1},b_{2},b_{3};-g_c) \right)$ where the large-order constraint $\sum a_i-\sum b_i -2=b_{\omega}$ has been employed and we found the result $\omega=0.85650$. This result can be compared with the recent Monte Carlo simulations calculations in Ref.\cite{nuN0} that predicts the result $\omega=\frac{\Delta_1}{\nu}=0.899(12)$ (see table-\ref{7L0} for comparison with different  methods). 

The field anomalous dimension is also given by:
\begin{equation}
\gamma_{\phi}\approx 0.05556g^{2}- 0.03704g^{3}+0.1929g^{4}-1.006g^{5}+7.095g^{6}- -57.74g^{7}.
\end{equation}
The suitable Hypergeometric  approximant used is 
\begin{equation}
\gamma_{\phi}=\, _4F_2\left(a_{1},a_{2},a_{3}%
,a_{4} ;b_{1},b_{2} ;-1\right)-\left(1+g\frac{ a_{1}a_{2}a_{3}a_{4} }{b_{1}b_{2}}\right).
\end{equation}
The  critical exponent $\eta$ is obtained from the relation $\eta=2\gamma_{\phi}(g_c)$ where we get the result $\eta=0.03129$. In a recent conformal
bootstrap calculation the result  $\eta=2\Delta_{\phi}-1=0.0282(4)$ has been obtained \cite{BstrabN0} while the Monte Carlo result is $\eta=0.031043(3)$ in Refs.\cite{nuN0E,ON17}.

For the mass anomalous dimension $\gamma_{m^2}$, the series up to seven-loops order is given by:
\begin{equation}
\gamma_{m^2}\approx -0.6667g+ 0.5556 g^{2}-2.056 g^{3}+10.76g^{4}-75.70 g^{5}+ 636.7g^{6}-6080g^{7}.\nonumber
\end{equation}  
The Hypergeometric approximant used is    $\left( _{5}F_{3}(a_{1},a_{2},a_{3}%
,a_{4},a_{5};b_{1},b_{2},b_{3};-g)-1\right)$ which corresponds to the Meijer-G function:

\begin{equation}
\gamma_{m^2}=\left(  \frac{\Gamma(b_{1})\Gamma(b_{2})\Gamma(b_{3})}{\Gamma\left(  a_{1}\right)
\Gamma\left(  a_{2}\right)  \Gamma\left(  a_{3}\right)  \Gamma\left(
a_{4}\right)  \Gamma\left(
a_{5}\right)} \MeijerG*{1}{5}{5}{4}{1-a_{1},1-a_{2},1-a_{3},1-a_{4},1-a_{5}}{0,1-b_{1}%
,1-b_{2},1-b_{3}}{-g }\right)-1.
\end{equation}
The critical exponent $\nu$ is then obtained as $\nu=\left(  2+\gamma_{m^2}\left(  g_{c}\right)  \right)  ^{-1}$ which yields the result $\nu=0.58723$.
This result can be compared with conformal bootstrap prediction $\nu=0.5877(12)$ in Ref.\cite{BstrabN0} and the Monte Carlo result $\nu=0.5875970(4)$ in Ref.\cite{nuN0}.
	\begin{table}[ht]
\caption{{\protect\scriptsize { The seven-loops  (7L)  Hypergeometric-Meijer
resummation  for the critical exponents $\nu,\eta$ and $\omega$  of the  self-avoiding walks model $(N=0)$. Here we compare with our results from previous section($\varepsilon^6$), conformal bootstrap (CB) calculations  \protect\cite{BstrabN0}, Monte Carlo simulation (MC) for $\nu$ from Ref.\protect\cite{nuN0E,ON17} and $\eta$ from Ref.\protect\cite{nuN0}. The six-loops Borel with conformal mapping (BCM) resummation ($\varepsilon^6$) from Ref.\protect\cite{ON17} and five-loops ($\varepsilon^5$) from same reference.}}}%
\label{7L0}
\begin{tabular}{|l|l|l|l| }
\hline
\ \ Method\ \ & $\ \ \ \ \ \  \nu$ & $\ \ \ \ \ \ \eta$ & $\ \ \ \ \ \omega$   \\ \hline
\ \ \begin{tabular}[c]{@{}l@{}}7L: This Work\\$\varepsilon^6$: This Work\\ \ \ \ CB\\  \ \ \ MC\\ $\varepsilon^6$: BCM \\ $\varepsilon^5$: BCM \end{tabular}\ \ & \begin{tabular}[c]{@{}l@{}}0.58723 \\0.58744\\ 0.5877(12)\\  0.5875970(4)\\ 0.5874(3)\\0.5873(13)\end{tabular} & \begin{tabular}[c]{@{}l@{}}0.03129 \\ 0.03034\\ 0.0282(4)\\  0.031043(3)\\ 0.0310(7)\\0.0314(11)\end{tabular} & \begin{tabular}[c]{@{}l@{}}0.85650 \\ 0.85559\\ \ \ \  \textbf{---}\\ 0.899(12)\\0.841(13)\\0.835(11)\end{tabular}  \\ \hline 

\end{tabular}%
\end{table}
\subsection{Resummation results for Ising universality class ( $N=1$)}

For $N=1,$ the seven-loops $\beta-$ function that has been recently obtained \cite{7L} is given by:
\begin{equation}
\beta\approx-\varepsilon g+3.000g^{2}-5.667g^{3}+32.55g^{4}-271.6g^{5}%
+2849g^{6}-34776g^{7}+474651g^{8}.
\end{equation}
The suitable approximant for this series is $\left( _{5}F_{3}(a_{1},a_{2},a_{3}%
,a_{4},a_{5};b_{1},b_{2},b_{3};-g)-1\right)$ which we used to obtain
 the critical coupling $g_c$ at which $\beta=0.$ In three dimensions ($\varepsilon=1$), the predicted critical
coupling is $g_{c}=0.47947$. This value can be compared with the five-loops resummation in table-\ref{gc345}. The critical exponent $\omega$ also predicted to have the value $0.82790$.  The conformal bootstrap calculation gives the result $\omega=0.8303(18)$ in Ref.\cite{Bstrab2} while Monte Carlo simulations result is $\omega=0.832(6)$ \cite{MC10}.
 
 The seven-loops perturbation series for the anomalous mass dimension $\gamma_{m^2}$ has been obtained in the same reference \cite{7L}
where:%
\begin{equation}
\gamma_{m^2}\approx-g+0.8333g^{2}-3.500g^{3}+19.96g^{4}-150.8g^{5}%
+1355g^{6}-13760g^7.
\end{equation}
We used  $\left(_{5}F_{3}(a_{1},a_{2},a_{3},a_{4},a_{5};b_{1},b_{2},b_{3};-g)-1\right)$  too for the resummation of this series. The $\nu$-exponent is then

	\[\
\nu=\left(  2+\gamma_{m^2}\left(  g_{c}\right)  \right)  ^{-1}=0.62934.
\]
The recent Monte Carlo prediction gives the value $\nu=0.63002(10)$ in
Ref.\cite{MC10} while in Ref.\cite{Bstrab2} one can find the result $\nu=0.62999(5)$ using conformal bootstrap calculations.
 
 The seven-loops order of the perturbation series for the   field anomalous dimension $\gamma_{\phi}$ is also   obtained  in Ref.\cite{7L} as:
\begin{equation}
\gamma_{\phi}\approx0.08333g^{2}-0.06250g^{3}+0.3385g^{4}-1.926g^{5}%
+14.38g^{6}-124.2g^7.
\end{equation}
We used the Hypergeometric approximant": 
	\begin{equation}\gamma\approx  {_4}F_{2}\left(a_{1},a_{2},a_{3}%
,a_{4} ;b_{1},b_{2};(-g)\right)- ( 1-\frac{a_{1}a_{2}a_{3}a_{4}}{b_{1}b_{2}}(-g)  )
\end{equation}
 to resum that series and the exponent $\eta$ is obtained from the relation
$\eta=2\gamma(g_c)$. We get the result $\eta=0.03684$. This result is compatible  with the recent conformal bootstrap calculation of  $\eta=0.03631(3)$ \cite{Bstrab2} and Monte Carlo simulation result of $\eta=0.03627(10)$ in Ref.\cite{MC10}. 
\begin{table}[ht]
\caption{{\protect\scriptsize { The seven-loops   Hypergeometric-Meijer
resummation  for the critical exponents $\nu,\eta$ and $\omega$  of the $O(1)$-symmetric model. Here we compare with our results from previous section($\varepsilon^6$), conformal bootstrap calculations  from Ref. \protect\cite{Bstrab2} and Monte Carlo simulation (MC)  from Ref.\protect\cite{MC10}. The six-loop Borel with conformal mapping (BCM) resummation ($\varepsilon^6$) from Ref.\protect\cite{ON17} and five-loops ($\varepsilon^5$) from same reference. The very recent calculations of critical exponents using nonperturbative renormalization group (NPRG)\protect\cite{NPRG} is listed last where results for $\nu$ and $\eta$ are up to $O(\partial^6)$ while for $\omega$ is up to $O(\partial^4$). }}}%
\label{7L1}
\begin{tabular}{|l|l|l|l| }
\hline
\ \ Method\ \ & $\ \ \ \ \ \  \nu$ & $\ \ \ \ \ \ \eta$ & $\ \ \ \ \ \omega$   \\ \hline
\ \ \begin{tabular}[c]{@{}l@{}}7L: This Work\\$\varepsilon^6$: This Work\\ \ \ \ CB\\  \ \ \ MC\\ $\varepsilon^6$: BCM \\ $\varepsilon^5$: BCM\\NPRG \end{tabular}\ \ & \begin{tabular}[c]{@{}l@{}}0.62934 \\0.62937\\ 0.62999(5)\\  0.63002(10)\\ 0.6292(5)\\0.6290(20)\\ 0.63012(16)\end{tabular} & \begin{tabular}[c]{@{}l@{}}0.03684 \\ 0.03545\\ 0.03631(3)\\  0.03627(10)\\ 0.0362(6)\\0.0366(11)\\0.0361(11) \end{tabular} & \begin{tabular}[c]{@{}l@{}}0.82790 \\ 0.82929\\ 0.8303(18)\\ 0.832(6)\\0.820(7)\\0.818(8)\\0.832(14)\end{tabular}  \\ \hline 

\end{tabular}%
\end{table}

\subsection{Resummation results for $N=2$ ($XY$ universality class)}
In this case, the seven-loops $\beta$-function is given by:

\begin{equation}
\beta\approx- g+3.333g^{2}-6.667g^{3}+39.95g^{4}-350.5g^{5}%
+3845g^{6}-48999g^{7}+696998g^{8}.
\end{equation}
This series is resummed using the approximant $\left(-g\left( _{5}F_{3}(a_{1},a_{2},a_{3},a_{4},a_{5};b_{1},b_{2},b_{3};-g\right)\right)$ which gives the  critical coupling value $g_c=0.43292$. Resuming  the g-differentiated series yields the result $\omega=0.80233$. The value $\omega=0.789$ has been adopted using a recent high-precision Monte Carlo calculations \cite{MC19} while the conformal bootstrap calculations gives $\omega=0.811(10)$ \cite{Bstrab3,ON17}

The mass anomalous dimension has the seventh loop result as: 
\begin{equation}
\gamma_{m^2}\approx- 1.333g+1.111g^{2}-5.222g^{3}+31.87g^{4}-255.8g^{5}%
+2434g^{6}-26086g^{7},
\end{equation}
where we resummed it using  $\left(_{5}F_{3}(a_{1},a_{2},a_{3},a_{4},a_{5};b_{1},b_{2},b_{3};-g)-1\right)$. This led to the result $\nu=0.66953$. The resent Monte Carlo result is $\nu=0.671 83(18)$ \cite{MC19} while the conformal bootstrap gives $\nu=0.6719(11)$ \cite{Bstrab4}. 

For the field anomalous dimension $\gamma_{\phi}$ we have:
\begin{equation}
\gamma_{\phi}\approx 0.11111g^{2}-0.09259g^{3}+0.5093g^{4}-3.148g^{5}%
+24.71g^{6}-224.6g^{7},
\end{equation}
The corresponding Hypergeometric approximant is  $0.11111g^{2}\left(_{4}F_{2}(a_{1},a_{2},a_{3},a_{4} ;b_{1},b_{2} ;-g)\right)$ with the result $\eta=0.03824$. For that exponent, the recent Monte Carlo simulations in Ref.\cite{MC19} gives $\eta=0.038 53(48)$ while conformal bootstrap gives the result $\eta=0.03852(64)$ \cite{Bstrab4}.
\begin{table}[ht]
\caption{{\protect\scriptsize { The seven-loops   Hypergeometric-Meijer
resummation  for the critical exponents $\nu,\eta$ and $\omega$  of the $O(2)$-symmetric model. For  comparison, other predictions are listed  from   previous section($\varepsilon^6$), conformal bootstrap calculations \protect\cite{Bstrab4} for $\nu$  and $\eta$, while $\omega$ from Ref. \protect\cite{Bstrab3,ON17}.  MC calculations  from Ref.\protect\cite{MC19}. The six-loop BCM resummation ($\varepsilon^6$) from Ref.\protect\cite{ON17} and five-loops ($\varepsilon^5$) from same reference while   NPRG results up to $O(\partial^4$) \protect\cite{NPRG} are listed last. }}}%
\label{7L2}
\begin{tabular}{|l|l|l|l| }
\hline
\ \ Method\ \ & $\ \ \ \ \ \  \nu$ & $\ \ \ \ \ \ \eta$ & $\ \ \ \ \ \omega$   \\ \hline
\ \ \begin{tabular}[c]{@{}l@{}}7L: This Work\\$\varepsilon^6$: This Work\\ \ \ \ CB\\  \ \ \ MC\\ $\varepsilon^6$: BCM \\ $\varepsilon^5$: BCM\\NPRG \end{tabular}\ \ & \begin{tabular}[c]{@{}l@{}}0.66953 \\0.66962\\ 0.6719(11)\\  0.67183(18)\\ 0.6690(10)\\0.6687(13)\\ 0.6716(6)\end{tabular} & \begin{tabular}[c]{@{}l@{}}0.03824 \\ 0.03733\\ 0.03852(64)\\  0.03853(48)\\ 0.0380(6)\\0.0384(10)\\0.0380(13) \end{tabular} & \begin{tabular}[c]{@{}l@{}}0.80233 \\ 0.80580\\ 0.811(10)\\ 0.789\\0.804(3)\\0.803(6)\\0.791(8)\end{tabular}  \\ \hline 
\end{tabular}%
\end{table}

\subsection{Resummation results  for Heisenberg universality class $(N=3)$}
The seven-loops $\beta$-function for $N=3$ is given by:

\begin{equation}
\beta\approx- g+3.667g^{2}-7.667g^{3}+47.65g^{4}-437.6g^{5}+4999g^{6}-66243g^{7}+978330g^{8}.
\end{equation}

 To resum this series, we  used the Hypergeometric approximant$\left(-g +  3.667 g^2 - 7.667 g^3(_{4}F_{2}(a_{1},a_{2},a_{3},a_{4};b_{1},b_{2};-g)\right)$ which predicts the critical coupling value  $g_c=0.39363$ while the resummation of the $\omega$-series gives the value  $0.78683$. Conformal bootstrap result is $\omega=0.791(22)$ \cite{Bstrab3,ON17} and the  Monte Carlo result is  $\omega=0.773$ \cite{MC01}.

The series representing the mass anomalous dimension up to seven-loop order is:
\begin{equation}
\gamma_{m^2}\approx- 1.667g+1.389g^{2}-7.222g^{3}+46.64g^{4}-394.9g^{5}%
+3950 ^{6}-44412g^{7},
\end{equation}
which has been resummed using $\left(_{5}F_{3}(a_{1},a_{2},a_{3},a_{4},a_{5};b_{1},b_{2},b_{3};-g)-1\right)$ that gives the result $\nu=0.70810$. In Ref.\cite{Bstrab4}, conformal bootstrap calculations gives the value $\nu=0.7121(28)$ and the  Monte Carlo simulations in Ref.\cite{MC11} gives $  \nu=0.7116(10)$.

The   field anomalous dimension $\gamma_{\phi}$ has  the seventh order perturbative form:
\begin{equation}
\gamma_{\phi}\approx 0.1389g^{2}-0.1273g^{3}+0.6993g^{4}-4.689g^{5}%
+38.44g^{6}-365.9g^{7},
\end{equation}
which approximated by  
 $ \left(g(_{4}F_{2}(a_{1},a_{2},a_{3},a_{4} ;b_{1},b_{2} ;-g)-1)\right)$ and gives the result $\eta=0.03795$. To compare with other recent results, the bootstrap calculations in Ref. \cite{Bstrab4} gives  $\eta=0.0386(12)$ and the Monte Carlo results gives $\eta=0.0378(3)$ \cite{MC11} . 
\begin{table}[ht]
\caption{{\protect\scriptsize { The seven-loops   Hypergeometric-Meijer
resummation  for the critical exponents $\nu$ , $\eta$ and $\omega$  of the $O(3)$-symmetric model. The results are  compared with our results from previous section ($\varepsilon^6$), conformal bootstrap calculations from Ref. \protect\cite{Bstrab4} for $\nu$  and $\eta$, while $\omega$ from Refs\protect\cite{Bstrab3,ON17}.  For MC simulations  $\omega$ is taken from from Ref.\protect\cite{MC01} while $\nu$ and $\eta$ are taken from from Ref.\protect\cite{MC11}. The six-loop BCM resummation   is taken from Ref.\protect\cite{ON17} and five-loops   from same reference. The very recent calculations NPRG \protect\cite{NPRG} is listed last and up to $O(\partial^4$). }}}%
\label{7L3}
\begin{tabular}{|l|l|l|l| }
\hline
\ \ Method\ \ & $\ \ \ \ \ \  \nu$ & $\ \ \ \ \ \ \eta$ & $\ \ \ \ \ \omega$   \\ \hline
\ \ \begin{tabular}[c]{@{}l@{}}7L: This Work\\$\varepsilon^6$: This Work\\ \ \ \ CB\\  \ \ \ MC\\ $\varepsilon^6$: BCM \\ $\varepsilon^5$: BCM\\NPRG \end{tabular}\ \ & \begin{tabular}[c]{@{}l@{}}0.70810 \\0.70722\\ 0.7121(28)\\  0.7116(10)\\ 0.7059(20)\\0.7056(16)\\ 0.7114(9)\end{tabular} & \begin{tabular}[c]{@{}l@{}}0.03795\\ 0.037301\\ 0.0386(12)\\  0.0378(3)\\ 0.0378(5)\\0.0382(10)\\0.0376(13) \end{tabular} & \begin{tabular}[c]{@{}l@{}}0.78683 \\ 0.79272\\ 0.791(22)\\ 0.773\\0.795(7)\\0.797(7)\\0.769(11)\end{tabular}  \\ \hline 

\end{tabular}%
\end{table}
\subsection{Resummation results for the  $O(4)$-symmetric case}
The seven-loops $\beta$-function for $N=4$  is shown to be:

\begin{equation}
\beta\approx- g+4.000g^{2}-8.667g^{3}+55.66g^{4}-533.0g^{5}%
+6318g^{6}-86768g^{7}+1.326 \times10^6g^{8}.
\end{equation}
The corresponding approximant is $\left(-g(_{5}F_{3}(a_{1},a_{2},a_{3},a_{4},a_{5} ;b_{1},b_{2},b_{3} ;-g)\right)$ which yields $g_c=0.36662$ while resumming the $\omega$-series gives the result  $\omega=0.80325 $. Monte Carlo Methods in Ref.\cite{MC01} gives $\omega=0.765$ while conformal bootstrap calculations predict the result $\omega=0.817(30)$ \cite{Bstrab3,ON17}.

The  anomalous mass dimension  is given by:

\begin{equation}
\gamma_{m^2}\approx- 2.000g+1.667g^{2}-9.500g^{3}+64.39g^{4}-571.9g^{5}+5983g ^{6}-70240g^{7},
\end{equation}
which has been approximated by  $_{5}F_{3}(a_{1},a_{2},a_{3},a_{4},a_{5};b_{1},b_{2},b_{3};-g)-1$  and gives $\nu=0.75093$. This result is very close to the Monte Carlo result $\nu=0.750(2)$ in Ref.\cite{MC11} and the conformal bootstrap result $\nu=0.751(3)$ in Ref.\cite{Bstrab3}.

Likewise, the    field anomalous dimension  up to seven loops is given by:
\begin{equation}
\gamma_{\phi}\approx 0.1667g^{2}-0.1667g^{3}+0.9028g^{4}-6.563g^{5}%
+55.93g^{6}-555.2g^{7},
\end{equation}
which is approximated by  
 $ g(_{4}F_{2}(a_{1},a_{2},a_{3},a_{4} ;b_{1},b_{2} ;-g)-1)$ and gives the result $\eta=0.03740$. Again the Monte Carlo simulations in Ref.\cite{MC11}   gives the values $\eta=0.0365(3)$. Also Monte Carlo simulations   and finite-size scaling of 3D Potts Models in  Ref.\cite{MC16} gives the result $\eta=5-2y_h=0.036(6)$ and the  conformal bootstrap calculations  is $0.0378(32)$ \cite{Bstrab5}. 

\begin{table}[ht]
\caption{{\protect\scriptsize { The seven-loops   Hypergeometric-Meijer
resummation  for the critical exponents $\nu$ , $\eta$ and $\omega$  of the $O(4)$-symmetric model. Here we compare with our results from previous section ($\varepsilon^6$), conformal bootstrap calculations \protect\cite{Bstrab3,ON17} for $\nu$  and $\omega$, while $\eta$ from Ref.\protect\cite{Bstrab5}. MC simulations  for  $\omega$ is taken from Ref.\protect\cite{MC01} while $\nu$ and $\eta$  are from Ref.\protect\cite{MC11}. The six-loop BCM resummation ($\varepsilon^6$) is taken from Ref.\protect\cite{ON17} and five-loops ($\varepsilon^5$) from same reference. NPRG results up to $O(\partial^4$) \protect\cite{NPRG} are shown in the last row. }}}%
\label{7L4}
\begin{tabular}{|l|l|l|l| }
\hline
\ \ Method\ \ & $\ \ \ \ \ \  \nu$ & $\ \ \ \ \ \ \eta$ & $\ \ \ \ \ \omega$   \\ \hline
\ \ \begin{tabular}[c]{@{}l@{}}7L: This Work\\$\varepsilon^6$: This Work\\ \ \ \ CB\\  \ \ \ MC\\ $\varepsilon^6$: BCM \\ $\varepsilon^5$: BCM\\NPRG \end{tabular}\ \ & \begin{tabular}[c]{@{}l@{}}0.750935 \\0.74151\\ 0.751(3)\\  0.750(2)\\ 0.7397(35)\\0.7389(24)\\ 0.7478(9)\end{tabular} & \begin{tabular}[c]{@{}l@{}}0.03740\\ 0.03621\\ 0.0378(32)\\  0.0360(3)\\ 0.0366(4)\\0.0370(9)\\0.0360(12) \end{tabular} & \begin{tabular}[c]{@{}l@{}}0.80325 \\ 0.76793\\ 0.817(30)\\ 0.765 (30)\\0.794(9)\\0.795(6)\\0.761(12)\end{tabular}  \\ \hline 

\end{tabular}%
\end{table}
A note to be mentioned is that one should not judge the convergence of the seven-loops resummation results by comparing with six-loops resummation  or lower order resummation in this work. The point is that the seven-loops resummation in this work applied for the $g$-series but for the other orders we resummed the $\varepsilon$-series. Our aim behind resumming both available series is to test our algorithm using  different types of perturbation series. To have an idea about the good convergence of our algorithm for the resummation of the $g$-series one should look at different orders of resummation of the $g$-series itself. For instance, for $N=4$, we get $\omega=0.77963$  from five-loop resummation of the $g$-series, $\omega=0.78162$ from six loops compared to the seven-loops  result in table-\ref{7L4} as $\omega=0.80325$.
\section{  Summary and Conclusions \label{conc}}
We show that divergent series with different large-order
behaviors can be approximated by different generalized Hypergeometric functions
$_{\text{ }p}F_{q}(a_{1},...a_{p};b_{1}....b_{q};\sigma z)$. The relation
between the number of numerator and denominator parameters ($p$ and $q$) is determined from the growth factor in
the large-order behavior of the divergent series. For a divergent series with
a growth  factor $n!$, the series expansion of the Hypergeometric function
$_{\text{ }p}F_{q}(a_{1},...a_{p};b_{1}....b_{q};\sigma z)$ where $p=q+2$ can
reproduce a large-order behavior with same growth  factor. Accordingly, the
Hypergeometric function $_{\text{ }p}F_{p-2}(a_{1},...a_{p};b_{1}%
....b_{p-2};\sigma z)$ is the suitable candidate to approximate such type of
divergent series. Since the function $_{\text{ }p}F_{p-2}(a_{1},...a_{p}%
;b_{1}....b_{-p-2};\sigma z)$ possesses an expansion of zero-radius of
convergence, a representation in terms of Meijer-G function is capable to
resum the divergent Hypergeometric series. 

 For divergent series that have growth factors $(2n)!$ and $(3n)!$, Hypergeometric functions with $p=q+3$ and
$p=q+4$, respectively, can reproduce such large order behaviors and thus are suitable approximants for such perturbation series. On the other hand, one might have a divergent series with finite radius of convergence  which has a large order behavior with a growth factor of $1$. To mimic such type of large order behavior,  the Hypergeometric function $_{\text{ }p}F_{p-1}(a_{1},...a_{p};b_{1}%
....b_{p-1};\sigma z)$ can be used   as suitable  approximant for such kind of divergent series.

The large-order behavior of the $\varepsilon$-expansion of the renormalization
group functions for the $O(N)$-symmetric model has a growth factor of  $n!$.
Accordingly, we used the Hypergeometric function $_{\text{ }p}F_{p-2}%
(a_{1},...a_{p};b_{1}....b_{p-2};\sigma z)$ to approximate the respective
divergent series. Since the strong-coupling data is not yet known for such
expansion, we use weak-coupling and large-order data to parametrize the
Hypergeometric function $_{\text{ }p}F_{p-2}(a_{1},...a_{p};b_{1}%
....b_{p-2};\sigma z)$. The parametrization of the Hypergeometric function is then followed by the resummation step of using a representation in terms of
Meijer-G function.  We applied the algorithm to resum the divergent series representing  critical exponents $\nu$ $(\nu^{-1})$, $\eta$ and $\omega$ as well as the critical coupling  up to $\varepsilon^{5}$ order as input. For $N$ equals $0,1,2,3$ and $4$, the results ought to be reasonable even for very low order of perturbation used to parametrize the Hypergeometric approximant. The results are greatly improved in using third order and being more precise in  going to fourth order while   the fifth order offers very competitive predictions when compared to other  resummation algorithms in literature.  

To show that the precise results extends to higher $N$ values, we resummed the perturbation series for the exponent $\nu$ for $N= 6,8,10$ and $12$. The precision of the results can be seen  from table-\ref{tab:5lresN} where we listed the    $5^{th}$ order  resummation results for the exponent $\nu$   and compared it with other methods.
 
All the Hypergeometric functions $_{\text{ }p}F_{p-2}(a_{1},...a_{p};b_{1}....b_{p-2};\sigma z)$   share the same analytic behavior. Accordingly, one expects no surprises  in going to higher orders of resummation. To test this clear fact as well as to seek more improved results, we resummed the six-loops order for the perturbation series for the exponents $\nu, \eta$ and $\omega$ for $N=1,2,3$ and $4$. The results are showing improved predictions  for those exponents. When compared to other calculations, our results for the critical exponents  are compatible  with the recent six-loops BC resummation  method in Ref\cite{ON17}, MC simulations calculations \cite{MC01,MC02,MC10,MC11,MC16,MCN2,MC19} and conformal bootstrap methods \cite{Bstrab,Bstrab,BstrabN0,Bstrab3,Bstrab4,Bstrab5}.
 
The very recent  seven-loops order (coupling-series) for the renormalization group functions $\beta, \gamma_{\phi}$ and $\gamma_{m^2}$  has been resummed too. Up to the best of our knowledge, no other resummation algorithm has been used to resum this order. Very accurate results for the critical coupling and the exponent $\nu$ have been extracted from the resummed functions. 
     
In all of our calculations, we used weak-coupling and large-order data as
input. The $a_{i}$ parameters in the Hypergeometric functions $_{\text{ }%
p}F_{q}(a_{1},...a_{p};b_{1}....b_{p-2};\sigma z)$ are well known to represent
the strong-coupling data \cite{Abo-large}. However, the strong coupling expansion for the
series under consideration has not been obtained yet ( up to the best of our
knowledge ). Accordingly, we cannot get benefited from this fact in further
acceleration of the convergence of the resummation algorithm. However,  the expansion coefficients of the Hypergeometric function depend on the strong-coupling parameters and they in turn constrained to mach the weak-coupling and large-order data. Accordingly,    this algorithm is linking the unknown strong-coupling  parameters  to the known weak-coupling and   large-order data. Thus the algorithm  has the ability  to predict the non-perturbative asymptotic strong-coupling  behavior of a quantum field theory from knowing  the weak coupling and large-order data. In other algorithms, this asymptotic behavior  is predicted from optimization techniques and different optimizations can even lead to different results for the same theory.
\section*{Acknowledgment}
We are very grateful to Oliver Schnetz for making the access to his maple package for the generation of  the seven-loops $g$-series available and for his valuable advice about how to  run the code.

\end{document}